\documentclass[sigconf]{acmart}
\usepackage[linesnumbered,ruled,vlined]{algorithm2e}
\usepackage{pifont}
\usepackage{float}
\usepackage{graphicx} 
\usepackage{tabularx}
\usepackage{enumitem}
\usepackage{bbding}
\usepackage{caption}
\usepackage{subcaption} 
\usepackage{amsmath}
\DeclareMathOperator*{\argmax}{arg\,max}
\DeclareMathOperator*{\argmin}{arg\,min}

\makeatletter
\pretocmd{\@chooseSymbol}{\raisebox{-.5ex}[\height][0pt]}{}{}
\makeatother

\AtBeginDocument{%
  }

\setcopyright{acmcopyright}
\copyrightyear{2018}
\acmYear{2018}
\acmDOI{XXXXXXX.XXXXXXX}

\acmConference[SIGMOD '24]{In Proceedings of the 2024 International Conference on
Management of Data}{June 08--15,
  2024}{Santiago, Chile}
\acmPrice{15.00}
\acmISBN{978-1-4503-XXXX-X/18/06}

\begin{document}
\sloppy
\title[RaBitQ: Quantizing High-Dim. Vectors with a Theoretical Error Bound for Approximate Nearest Neighbor Search]{RaBitQ: Quantizing High-Dimensional Vectors with a Theoretical Error Bound for Approximate Nearest Neighbor Search}

\author{Jianyang Gao}
\email{jianyang.gao@ntu.edu.sg}
\orcid{0009-0008-4684-3624}
\affiliation{%
  \institution{Nanyang Technological University}
  \country{Singapore}
}

\author{Cheng Long}
\authornote{Cheng Long is the corresponding author.}
\email{c.long@ntu.edu.sg}
\orcid{0000-0001-6806-8405}
\affiliation{%
  \institution{Nanyang Technological University}
  \country{Singapore}
}

\renewcommand{\shortauthors}{Jianyang Gao and Cheng Long}

\begin{abstract}
Searching for approximate nearest neighbors (ANN) in the high-dimensional Euclidean space is a pivotal problem. Recently, with the help of fast SIMD-based implementations, Product Quantization (PQ) and its variants can often efficiently and accurately estimate the distances between the vectors and have achieved great success in the in-memory ANN search. Despite their empirical success, we note that these methods do not have a theoretical error bound and are observed to fail disastrously on some real-world datasets. Motivated by this, we propose a new randomized quantization method named RaBitQ, which quantizes $D$-dimensional vectors into $D$-bit strings. RaBitQ guarantees a sharp theoretical error bound and provides good empirical accuracy at the same time. In addition, we introduce efficient implementations of RaBitQ, supporting to estimate the distances with bitwise operations or SIMD-based operations. Extensive experiments on real-world datasets confirm that (1) our method outperforms PQ and its variants in terms of accuracy-efficiency trade-off by a clear margin and (2) its empirical performance is well-aligned with our theoretical analysis.
\end{abstract}

\maketitle

\section{Introduction}
\label{sec:introduction}

Searching for the nearest neighbor (NN) in the high-dimensional Euclidean space is pivotal for various applications such as information retrieval~\cite{KNNimageretrieval}, data mining~\cite{KNNClassficiation}, and recommendations~\cite{KNNrecommendation}. However, the curse of dimensionality~\cite{indyk1998approximate, vafile} makes exact NN queries on extensive vector databases practically infeasible due to their long response time. To strike a balance between time and accuracy, researchers often explore its relaxed counterpart, known as approximate nearest neighbor (ANN) search~\cite{datar2004locality, muja2014scalable, jegou2010product, malkov2018efficient, ge2013optimized, scann}.

Product Quantization (PQ) and its variants are a family of popular methods for ANN~\cite{jegou2010product,ge2013optimized,ITQ,additivePQ,lsq,lsq++,composite_quantization, learningtohash,learningtohashsurvey,ite_matsui_2018_pq_survey}. 
These methods target to efficiently estimate the distances between the data vectors and query vectors during the query phase to shortlist a list of candidates, which would then be re-ranked based on exact distances for finding the NN.
Specifically, during the \underline{index phase}, they (1) construct a quantization codebook and (2) find for each data vector the nearest vector in the codebook as its quantized vector. 
The quantized vector is represented and stored as a short quantization code (e.g., the ID of the quantized data vector in the codebook). 
During the \underline{query phase}, they (1) pre-compute the (squared) distances~\footnote{By distances, we refer to the squared distances without further specification.} between the query and the vectors in the codebook when a query comes and (2) for a data vector, they adopt the distances between the query vector and its quantized data vector (which can be computed by looking up the pre-computed values) as the estimated distances.
Recently, with the help of the fast SIMD-based implementation~\cite{fastscan,fastscanavx2}, 
PQ has achieved great success in the in-memory ANN search~\cite{NGT-QG,johnson2019billion_faiss,scann}. 
In particular, on many real-world datasets, the method can efficiently estimate the distances with high accuracy.

Despite their empirical success on many real-world datasets, to the best of our knowledge, none of PQ and its variants~\cite{jegou2010product,ge2013optimized,ITQ,additivePQ,lsq,lsq++,composite_quantization, learningtohash,learningtohashsurvey,ite_matsui_2018_pq_survey} provide theoretical error bounds on the estimated distances.
This is because they lose guarantees in both (1) the codebook construction component and (2) the distance estimation component. \underline{Codebook Construction}: 
They construct the codebook often via approximately solving an optimization problem for a heuristic objective function, e.g., PQ conducts KMeans clustering on the sub-segments of the data vectors and uses the set of the products of cluster centroids as the codebook. 
However, due to their heuristic nature, it is often difficult to analyze their results theoretically (e.g.,  no theoretical results have been achieved on the distance between a data vector and its nearest vector in the codebook (i.e., its quantized vector)).
\underline{Distance Estimation}: They estimate the distance between a data vector and a query vector with that between the quantized data vector and the query vector, i.e., they simply treat the quantized data vector as the data vector for computing the distance. While this looks intuitive, it does not come with a theoretical error bound on the approximation. The lack of a theoretical error bound indicates that these methods may unpredictably fail anytime, moderately or severely, when they are deployed in real-world systems to handle new datasets and queries which they have not been tested on. 
In fact, such failure has been observed on public real-world datasets which are widely adopted to benchmark ANN search.
For example, on the dataset MSong, PQ (with the fast SIMD-based implementation~\cite{fastscan,fastscanavx2}) incurs more than 50\% of average relative error on the estimated distances between the query and data vectors, which causes disastrous recall of ANN search (e.g., it has no more than 60\% recall even with re-ranking applied, as shown in Section~\ref{subsubsec: time-accuracy trade-off ann search}).

\begin{table*}[th]
\caption{Comparison between RaBitQ and PQ and its variants. More {\protect\FiveStar's} indicates better efficiency.}
\label{tab:comparison}
\vspace*{-4mm}
\begin{tabularx}{\linewidth}{|>{\hsize=.22\hsize}X|>{\hsize=.39\hsize}X|>{\hsize=.39\hsize}X|}
\hline
                            & RaBitQ (new)                                                    & PQ and its variants                                                                                                           \\ \hline
Codebook                    & Randomly transformed bi-valued vectors.                           & Cartesian product of sub-codebooks.                                                            \\ \hline
Quantization Code           & A bit string.                                                 & A sequence of 4-bit/8-bit unsigned integers.           \\ \hline
Distance Estimator                   & Unbiased and provides a sharp error bound. & Biased and provide no error bound. \\ \hline
Implementation (single)  & Bitwise operations. \FiveStar\FiveStar                                      & Looking up tables in RAM. \FiveStar  \\ \hline
Implementation (batch) & Fast SIMD-based operations. \FiveStar\FiveStar\FiveStar                             & Fast SIMD-based operations. \FiveStar\FiveStar \FiveStar                                                                                  \\ \hline
\end{tabularx}
\end{table*}

In this paper, we propose a new quantization method, which provides unbiased estimation on the distances and achieves a sharp~\footnote{The error bound is sharp in the sense that it achieves the asymptotic optimality shown in \cite{2017_focs_additive_error}.
Detailed discussions can be found in Section~\ref{subsubsec: construct unbiased estimator}.} theoretical error bound.
The new method achieves this with careful and integrated design in both the codebook construction and distance estimation components.  
\underline{Codebook Construction}: 
It first normalizes the data vectors in order to align them on the unit hypersphere in the $D$-dimensional space.
It then constructs the codebook by 
(1) constructing a set of $2^D$ \emph{bi-valued} vectors whose coordinates are $-1/\sqrt {D}$ or $+1/\sqrt {D}$ 
(i.e., the set consists of the vertices of a hypercube, which evenly spread on the unit hypersphere) and 
(2) randomly rotating the bi-valued vectors by multiplying each with a \emph{random orthogonal matrix}~\footnote{We note that we do not explicitly materialize the codebook, but maintain it conceptually, as existing quantization methods such as PQ do.} (i.e., it performs a type of Johnson-Lindenstrauss Transformation~\cite{johnson1984extensions}, JLT in short).
For each data vector, its nearest vector from the codebook is taken as the quantized vector.
Since each quantized vector is a rotated $D$-dimensional bi-valued vector, we represent its quantization code as a bit string of length $D$, where 0 and 1 indicate the two distinct values. 
The rationale of the codebook construction is that it has a clear geometric interpretation (i.e., the vectors in the codebook are a set of randomly rotated vectors on the unit hypersphere) 
such that it is possible to analyze the geometric relationship among the data vectors, their quantized vectors and the query vectors explicitly. 
\underline{Distance Estimation}: We carefully design an estimator of the distance between a data vector and a query vector by leveraging the aforementioned geometric relationship.
We prove that the estimator is \emph{unbiased} and has a sharp probabilistic error bound with the help of plentiful theoretical tools about the JLT~\cite{jlintro,vershynin_2018,uniform_spherical_distribution}.
This is in contrast to PQ and its variants, which simply treat the quantized vector as the data vector for estimating the distances, which is \emph{biased} and provides \emph{no} theoretical error bound. We call the new quantization method, which uses \underline{ra}ndomly transformed \underline{bi}-valued vec\underline{t}ors for \underline{q}uantizing data vectors, RaBitQ.
Compared with PQ and its variants, RaBitQ has its superiority not only in providing error bounds in theory, but also in estimating the distances with smaller empirical errors even with shorter quantization codes by roughly a half (as verified on all the tested datasets shown in Section~\ref{subsubsec: time-accuracy trade-off per vector}).

We further introduce two efficient implementations for computing the value of RaBitQ's distance estimator, namely one for a \emph{single} data vector and the other for a \emph{batch} of data vectors. For the former, our implementation is based on simple bitwise operations - recall that our quantization codes are bit strings. Our implementation is on average 3x faster than the original implementation of PQ which relies on looking up tables in RAM while reaching the same accuracy (as shown in Section~\ref{subsubsec: time-accuracy trade-off per vector}). Note that for a single data vector, the SIMD-based implementation of PQ~\cite{fastscan, fastscanavx2} is not feasible as it requires to pack the quantization codes in a batch and reorganize their layout carefully. For the latter, the same strategy of the fast SIMD-based implementation~\cite{fastscan, fastscanavx2} can be adopted seamlessly, and thus it achieves similar efficiency as existing SIMD-based implementation of PQ does when similar length quantization codes are used - in this case, our method would provide more accurate estimated distances as explained earlier. Table~\ref{tab:comparison} provides some comparison between RaBitQ and PQ and its variants.

We summarize our major contributions as follows.
\begin{enumerate}
\item We propose a new quantization method, namely RaBitQ. (1) It constructs the codebook via randomly transforming bi-valued vectors.
(2) It designs an \emph{unbiased} distance estimator with 
a sharp probabilistic error bound.

\item We introduce efficient implementations of computing the distance estimator for RaBitQ. Our implementation is more efficient than its counterpart of PQ and its variants when estimating the distance for a single data vector and is comparably fast when estimating the distances for a batch of data vectors with quantization codes of similar lengths.

\item We conduct extensive experiments on real-world datasets, which show that (1) RaBitQ provides more accurate estimated distances than PQ (and its variants) even when the former uses shorter codes than the latter by roughly a half (which implies the accuracy gap would be further larger when both methods use codes of similar lengths); (2) RaBitQ works stably well on all datasets tested including some on which PQ (and its variants) fail (which is well aligned with the theoretical results); (3) RaBitQ is superior over PQ (and its variants) in terms of time-accuracy trade-offs for in-memory ANN by a clear margin on all datasets tested; and (4) RaBitQ has its empirical performance well-aligned with the theoretical analysis.
\end{enumerate}

The remainder of the paper is organized as follows. 
Section~\ref{sec:preliminaries} introduces the ANN search and PQ and its variants.
Section~\ref{sec:RaBitQ method} presents our RaBitQ method.
Section~\ref{sec: RaBitQ for ANN} illustrates the application of RaBitQ to the in-memory ANN search.
Section~\ref{sec:experiments} provides extensive experimental studies on real-world datasets.
Section~\ref{sec:related work} discusses related work.
Section~\ref{sec:conclusion and discussion} presents the conclusion and discussion.
\section{ANN Query and Quantization}
\label{sec:preliminaries}

\smallskip \noindent \textbf{ANN Query.}
Suppose that we have a database of $N$ data vectors in the $D$-dimensional Euclidean space. 
The approximate nearest neighbor (ANN) search query is to retrieve the nearest vector from the database for a given query vector $\mathbf{q}$.
The question is usually extended to the query of retrieving the $K$ nearest neighbors. 
For the ease of narrative, we assume that $K=1$ in our algorithm description, while all of the proposed techniques can be easily adapted to a general $K$. 
We focus on the in-memory ANN, which assumes that all the raw data vectors and indexes can be hosted in the main memory~\cite{annbenchmark, malkov2018efficient, fu2019fast, li2019approximate, fastscan, fastscanavx2, adsampling}. 

\smallskip \noindent \textbf{Product Quantization.}
\label{subsubsec: product quantization}
Product Quantization (PQ) and its variants are a family of popular methods for ANN~\cite{jegou2010product,ge2013optimized,ITQ,additivePQ,lsq,lsq++,composite_quantization, learningtohash,learningtohashsurvey,ite_matsui_2018_pq_survey} (for the discussion on a broader range of quantization methods, see Section~\ref{sec:related work}).
For a query vector and a data vector, these methods target to efficiently estimate their distance based on some pre-computed short quantization codes.
Specifically, for PQ, it splits the $D$-dimensional vectors into $M$ sub-segments (each sub-segment has $D/M$ dimensions). 
For each sub-segment, it performs KMeans clustering on the $D/M$-dimensional vectors to obtain $2^k$ clusters and then takes the centroids of the clusters as a sub-codebook where $k$ is a tunable parameter which controls the size of the sub-codebook ($k=8$ by default).
The codebook of PQ is then formed by the Cartesian product of the sub-codebooks of the sub-segments and thus has the size of $(2^k)^M$. 
Correspondingly each quantization code can be represented as an $M$-sized sequence of $k$-bit unsigned integers. 
During the query phase, asymmetric distance computation is adopted to 
estimate the distance~\cite{jegou2010product}.
In particular, it pre-processes $M$ look-up-tables (LUTs) for each sub-codebook when a query comes. The $i$th LUT contains $2^k$ numbers which represent the squared distances between the vectors in the $i$th sub-codebook and $i$th sub-segment of the query vector. 
For a given quantization code, by looking up and accumulating the values in the LUTs for $M$ times, PQ can compute an estimated distance.

Recently, \cite{fastscan, fastscanavx2} propose a SIMD-based fast implementation for PQ (PQ Fast Scan, PQx4fs in short). They speed up the look-up and accumulation operations significantly, making PQ an important component in many popular libraries for in-memory ANN search such as Faiss from Meta~\cite{johnson2019billion_faiss}, ScaNN from Google~\cite{scann} and NGT-QG from Yahoo Japan~\cite{NGT-QG}.
At its core, unlike the original implementation of PQ which relies on looking up the LUTs in RAM, \cite{fastscan,fastscanavx2} propose to host the LUTs in SIMD registers and look up the LUTs with the SIMD shuffle instructions. 
To achieve so, the method makes several modifications on PQ. 
First, in order to fit the LUTs into the AVX2 256-bit registers, it modifies the original setting of $k=8$ to $k=4$ so that in each LUT, there are only $2^4$ floating-point numbers. It further quantizes the numbers in the LUT to be 8-bit unsigned integers so that one LUT takes the space of only 128 ($2^4 \times 8$) bits. Thus, one AVX2 256-bit register is able to host two LUTs. 
Second, in order to look up the LUTs efficiently, the method packs every 32 quantization codes in a batch and reorganizes their layout. 
In this case, a series of operations can estimate the distances for 32 data vectors all at once. 
Without further specification, by PQ, we refer to PQx4fs by default because without the fast SIMD-based implementation, the efficiency of PQ is much less competitive in the in-memory ANN search~\cite{fastscan, fastscanavx2} (see Section~\ref{subsubsec: time-accuracy trade-off per vector}). 

Nevertheless, none of PQ and its variants provide a theoretical error bound on the errors of the estimated distances~\cite{jegou2010product,ge2013optimized,ITQ,additivePQ,lsq,lsq++,composite_quantization, learningtohash,learningtohashsurvey,ite_matsui_2018_pq_survey}, as explained in Section~\ref{sec:introduction}. Indeed, we find that the accuracy of PQ can be disastrous (see Section~\ref{subsubsec: time-accuracy trade-off ann search}), e.g., on the dataset MSong, PQ cannot achieve $\ge 60\%$ recall even with re-ranking applied.
We note that Locality Sensitive Hashing (LSH) is a family of methods which promise rigorous theoretical guarantee ~\cite{indyk1998approximate,datar2004locality,sun2014srs,tao2010efficient,huang2015query,dblsh}. However, as is widely reported~\cite{learningtohashsurvey,annbenchmark,li2019approximate}, these methods can hardly produce competitive empirical performance.
Furthermore, their guarantees are on the accuracy of $c$-approximate NN query.
In particular, LSH guarantees to return a data vector whose distance from the query is at most $(1+c)$ times of a fixed radius $r$ with high probability (if there exists a data vector whose distance from the query is within the radius $r$). Due to the relaxation factor $c$, there can be many that satisfy the statement. 
The guarantee of returning \textit{any} of them does not help to produce high recall for ANN search. In contrast, a guarantee on the distance estimation can help to decide whether a data vector should be re-ranked for achieving high recall (see Section~\ref{sec: RaBitQ for ANN}).

\begin{table}
  \caption{Notations.}
  \label{tab:freq}
  \vspace{-4mm}
  \begin{tabularx}{\linewidth}{cX}
    \toprule
    Notation & Definition\\
    \midrule
    $\mathbf{o}_r, \mathbf{q}_r$  &  The raw data and query vectors.\\
    $\mathbf{o}, \mathbf{q}$   &  The normalized data and query vectors.\\
    $\mathcal{C}, \mathcal{C}_{rand}$    &  The quantization codebook, its randomized version. \\
    $P$    &  A random orthogonal transformation matrix. \\
    $\mathbf{\bar x}$ & The code in $\mathcal{C}$ s.t. $P \mathbf{\bar x}$ is the quantized vector of $\mathbf{o}$.\\
    $\mathbf{\bar o}$ & The quantized vector of $\mathbf{o}$ in $\mathcal{C}_{rand}$, i.e., $\mathbf{\bar o} = P \mathbf{\bar x}$. \\     
    $\mathbf{\bar x}_b$ &  
    The quantization code of $\mathbf{o}$ as a $D$-bit string.\\
    $\mathbf{q'}$ &  The inversely transformed query vector, i.e., $P^{-1} \mathbf{q}$.  \\
    $\mathbf{\bar q}$ &  The quantized query vector of $\mathbf{q'}$.  \\
    $\mathbf{\bar q}_u$ &  The unsigned integer representation of $\mathbf{\bar q}$.  \\
  \bottomrule
\end{tabularx}
\vspace{-6mm}
\end{table}

\section{The RaBitQ Method}
\label{sec:RaBitQ method}
In this section, we present the details of RaBitQ.
In Section~\ref{subsec: index}, we present the index phase of RaBitQ, which normalizes the data vectors (Section~\ref{subsubsec: pre-procession}), constructs a codebook (Section~\ref{subsubsec: the RaBitQ}) and computes the quantized vectors of data vectors (Section~\ref{subsubsec: quantization code}). 
In Section~\ref{subsec: unbiased estimator}, we introduce the distance estimator of RaBitQ, which is unbiased and provides a rigorous theoretical error bound.
In Section~\ref{subsec: efficient computation}, we illustrate how to efficiently compute the value of the estimator. 
In Section~\ref{subsec: summary}, we summarize the RaBitQ method.
Table~\ref{tab:freq} lists the frequently used notations and their definitions.

\subsection{Quantizing the Data Vectors with RaBitQ}

\label{subsec: index}
\subsubsection{Converting the Raw Vectors into Unit Vectors via Normalization}
\label{subsubsec: pre-procession}
We note that directly constructing a codebook for the raw data vectors is challenging for achieving the theoretical error bound because the Euclidean space is \textit{unbounded} and the raw data vectors may appear anywhere in the infinitely large space. 
To deal with this issue, a natural idea is to normalize the raw vectors into \emph{unit} vectors.
Specifically, let $\mathbf{c} $ be the centroid of the raw data vectors. 
We normalize the raw data vectors $\mathbf{o}_r $ to be $ \mathbf{o}:= \frac{\mathbf{o}_r- \mathbf{c} }{\| \mathbf{o}_r-\mathbf{c} \|}$. Similarly, we normalize the raw query vector $\mathbf{q}_r$ (when it comes in the query phase) to be  $\mathbf{q}:= \frac{\mathbf{q}_r-\mathbf{c} }{\| \mathbf{q}_r-\mathbf{c} \|}$.
The following expressions bridge 
the distance between the raw vectors (i.e., our target) and the inner product of the normalized vectors.
\begin{align}
    &\| \mathbf{o}_r-\mathbf{q}_r \|^2 
    = \| {(\mathbf{o}_r - \mathbf{c} )-( \mathbf{q}_r -\mathbf{c} )} \| ^2 
    \\= &\| {\mathbf{o}_r -\mathbf{c} }\|^2 + \| \mathbf{q}_r- \mathbf{c} \|^2 -2 \cdot \| \mathbf{o}_r-\mathbf{c} \| \cdot \| \mathbf{q}_r-\mathbf{c} \|  \cdot \left< \mathbf{q}, \mathbf{o} \right>   \label{eq:dis to ip}
\end{align}
We note that $\| \mathbf{o}_r-\mathbf{c}  \| $ is the distance from the data vector to the centroid, which can be pre-computed during the index phase. $\| \mathbf{q}_r-\mathbf{c}  \| $ is the distance from the query vector to the centroid. It can be computed during the query phase and its cost can be shared by all the data vectors. 
Thus, based on Equation (\ref{eq:dis to ip}), the question of computing $\| \mathbf{o}_r-\mathbf{q}_r \|^2 $ is reduced to that of computing the inner product of two unit vectors $\left< \mathbf{q}, \mathbf{o}  \right> $.
We note that in practice we can cluster the data vectors first (e.g., via KMeans clustering) and perform the normalization for data vectors within a cluster individually based on the centroid of the cluster. When considering the data vectors within a cluster, we normalize the query vector based on the corresponding centroid. In this way, the normalized data vectors are expected to spread evenly on the unit hypersphere, removing the skewness of the data (if any) to some extent.
For the sake of convenience, in the following parts without further clarification, by the data and query vector, we refer to their corresponding unit vectors.
With this conversion, 
we next focus on estimating the inner product of the unit vectors, i.e., $\left< \mathbf{q}, \mathbf{o}  \right> $. 

\subsubsection{ Constructing the Codebook}
\label{subsubsec: the RaBitQ}
As mentioned in Section~\ref{subsubsec: pre-procession}, the data vectors are supposed, to some extent, to be evenly spreading on the unit hypersphere due to the normalization. By intuition, our codebook should also spread evenly on the unit hypersphere.
To this end, a natural construction of the codebook is given as follows. 
\begin{align}
    \mathcal{C}:= \left\{ + \frac{1}{\sqrt {D} }, - \frac{1}{\sqrt {D} }  \right\}^D
\end{align}
It is easy to verify that the vectors in $\mathcal{C}$ are unit vectors and the codebook has the size of $|\mathcal{C}|=2^D$. 

However, such construction may favor some certain vectors and perform poorly for others. 
For example, for the data vector $(1 / \sqrt {D}, ..., 1 / \sqrt {D} )$, its quantized data vector (which corresponds to the vector in $\mathcal{C}$ closest from the data vector) is $(1 / \sqrt {D}, ..., 1 / \sqrt {D} )$, and its squared distance to the quantized data vector is $0$. 
In contrast, for the vector $(1,0,...,0)$, its quantized data vector is also $(1 / \sqrt {D}, ..., 1 / \sqrt {D} )$, and its squared distance to the quantized data vector equals to $2 - 2/\sqrt{D}$.
To deal with this issue, we inject the codebook some randomness. 
Specifically, let $P$ be a random orthogonal matrix.
We propose to apply the transformation $P$ to the codebook (which is one type of the Johnson-Lindenstrauss Transformation~\cite{johnson1984extensions}). 
Our final codebook is given as follows.
\begin{align}
    \mathcal{C}_{rand} := \left\{ P \mathbf{x} | \mathbf{x}  \in \mathcal{C}  \right\}  
\end{align}
Geometrically, the transformation simply rotates the codebook because the matrix $P$ is orthogonal, and thus, the vectors in $\mathcal{C}_{rand}$ are still unit vectors.
Moreover, the rotation is uniformly sampled from ``all the possible rotations'' of the space. 
Thus, for a unit vector in the codebook $\mathcal{C}$, it has equal probability to be rotated to anywhere on the unit hypersphere.
This step thus removes the preference of the deterministic codebook $\mathcal{C}$ on specific vectors.

We note that to construct the codebook $\mathcal{C}_{rand} $, we only need to sample a random transformation matrix $P$. To store the codebook $\mathcal{C}_{rand} $, we only need to physically store the sampled $P$ but not all the transformed vectors. 
The codebook constructed by this operation is much simpler than its counterpart in PQ and its variants which rely on approximately solving an optimization problem.

\subsubsection{Computing the Quantized Codes of Data Vectors}
\label{subsubsec: quantization code}
With the constructed codebook, the next step is to find the nearest vector from $\mathcal{C}_{rand}$ for each data vector as its quantized vector. 
For a \textit{unit} vector $\mathbf{o} $, to find its nearest vector, it is equivalent to find the one which has the largest inner product with it. Let $P\mathbf{\bar x} \in \mathcal{C}_{rand}$ be the quantized data vector (where $\mathbf{\bar x}\in \mathcal{C}$). The following equations illustrate the idea rigorously. 
\begin{align}
    \mathbf{\bar x} = & \argmin_{\mathbf{x} \in \mathcal{C}  } \| \mathbf{o} - P\mathbf{x}\|^2 \label{eq: nearest quantized vector}
    \\=&  \argmin_{\mathbf{x} \in \mathcal{C}  } (\| \mathbf{o}\|^2 + \| P\mathbf{x} \|^2  -  2 \left< \mathbf{o}, P\mathbf{x}  \right> )
    \label{eq: to inner product 1}
    \\=& \argmin_{\mathbf{x} \in \mathcal{C}  }  (2  -  2 \left< \mathbf{o}, P\mathbf{x}  \right>) =\argmax_{\mathbf{x} \in \mathcal{C}  } \left< \mathbf{o}, P\mathbf{x} \right>
    \label{eq: to inner product 2}
\end{align}
Equation (\ref{eq: nearest quantized vector}) is based on the definition of the quantized data vector.
Equation (\ref{eq: to inner product 1}) is due to elementary linear algebra operations. 
Equation (\ref{eq: to inner product 2}) is because $P \mathbf{x} $ and $\mathbf{o} $ are unit vectors. 
However, by Equation (\ref{eq: to inner product 2}), it is costly to find the quantized data vector by physically transforming the huge codebook and finding the nearest vector via enumeration. We note that the inner product is invariant to orthogonal transformation (i.e., rotation). Thus, instead of transforming the huge codebook, we \textit{inversely} transform the data vector $\mathbf{o}$. The following expressions formally present the idea.
\begin{align}
    \left< \mathbf{o}, P\mathbf{x} \right>=  \left< P^{-1}\mathbf{o}, P^{-1} P\mathbf{x} \right>  = \left< P^{-1}\mathbf{o}, \mathbf{x} \right>
\end{align}
Recall that the entries of $\mathbf{x} \in \mathcal{C}  $ are $\pm 1/\sqrt {D}$. To maximize the inner product, we only need to pick the $\mathbf{\bar x}\in \mathcal{C} $ whose signs of the entries match those of $P^{-1} \mathbf{o} $. Then $P\mathbf{\bar x}$ is the quantized data vector.

In summary, to find the nearest vector of a data vector $\mathbf{o} $ from $\mathcal{C}_{rand}$, we can inversely transform $\mathbf{o}$ with $P^{-1}$ and store the signs of its entries as a $D$-bit string $\mathbf{\bar x}_b \in \left\{ 0,1 \right\}^D $.
We call the stored binary string $\mathbf{\bar x}_b$ as the \emph{quantization code}, which can be used to re-construct the quantized vector $\mathbf{\bar x}$. 
Let $\mathbf{1}_D$ be the $D$-dimensional vector which has all its entries being ones.
The relationship between $\mathbf{\bar x}_b$ and $\mathbf{\bar x} $ is given as $\mathbf{\bar x} = (2\mathbf{\bar x}_b-\mathbf{1}_D )/\sqrt{D}$, i.e., when the $i$th coordinate $\mathbf{\bar x}_b[i] = 1$, we have $\mathbf{\bar x}[i] = 1/\sqrt{D}$ and when $\mathbf{\bar x}_b[i] = 0$, we have $\mathbf{\bar x}[i] = -1/\sqrt{D}$.
For the sake of convenience, we denote the quantized data vector as $\mathbf{\bar o}:= P \mathbf{\bar x}$. 

Till now, we have finished the pre-processing in the index phase. 
We note that the time cost in the index phase is not a bottleneck for our method, which is the same as in the cases of PQ and OPQ (a popular variant of PQ)~\cite{ge2013optimized}.
For example, on the dataset GIST with one million 960-dimensional vectors, with 32 threads on CPU, our method, PQ and OPQ take 117s, 105s and 291s respectively.
The space complexity of the methods is not a bottleneck for the in-memory ANN either, because the space consumption is largely due to the space for storing the raw vectors. 
As a comparison, each raw vector takes $32D$ bits (i.e., $D$ floating-point numbers). Our method by default has $D$ bits for a quantization code. PQ and OPQ by default have $2D$ bits for a quantization code (i.e., $M=D/2$) according to \cite{scann,faiss_github}, which is significantly smaller than the space for storing the raw vectors.

\subsection{Constructing an Unbiased Estimator}
\label{subsec: unbiased estimator}
Recall that the problem of computing $\| \mathbf{o}_r-\mathbf{q}_r \|^2 $ can be reduced to that of computing the inner product of two unit vectors $\left< \mathbf{o}, \mathbf{q}  \right> $.
In this section, we introduce an unbiased estimator for $\left< \mathbf{o,q} \right> $. 
Unlike PQ and its variants which simply treat the quantized data vector as the data vector for estimating the distances without theoretical error bounds, we first explicitly derive the relationship between $\left< \mathbf{o}, \mathbf{q}  \right>$ and $\left< \mathbf{\bar o}, \mathbf{q}  \right>$ in Section~\ref{subsubsec: geometric}. 
We then construct an unbiased estimator for $\left< \mathbf{o,q} \right> $ based on the derived relationships and present its rigorous error bound in Section~\ref{subsubsec: construct unbiased estimator}. 

\begin{figure}[thb]
    \centering
    \begin{subfigure}[b]{0.62\linewidth}
        \centering
        \includegraphics[width=\textwidth]{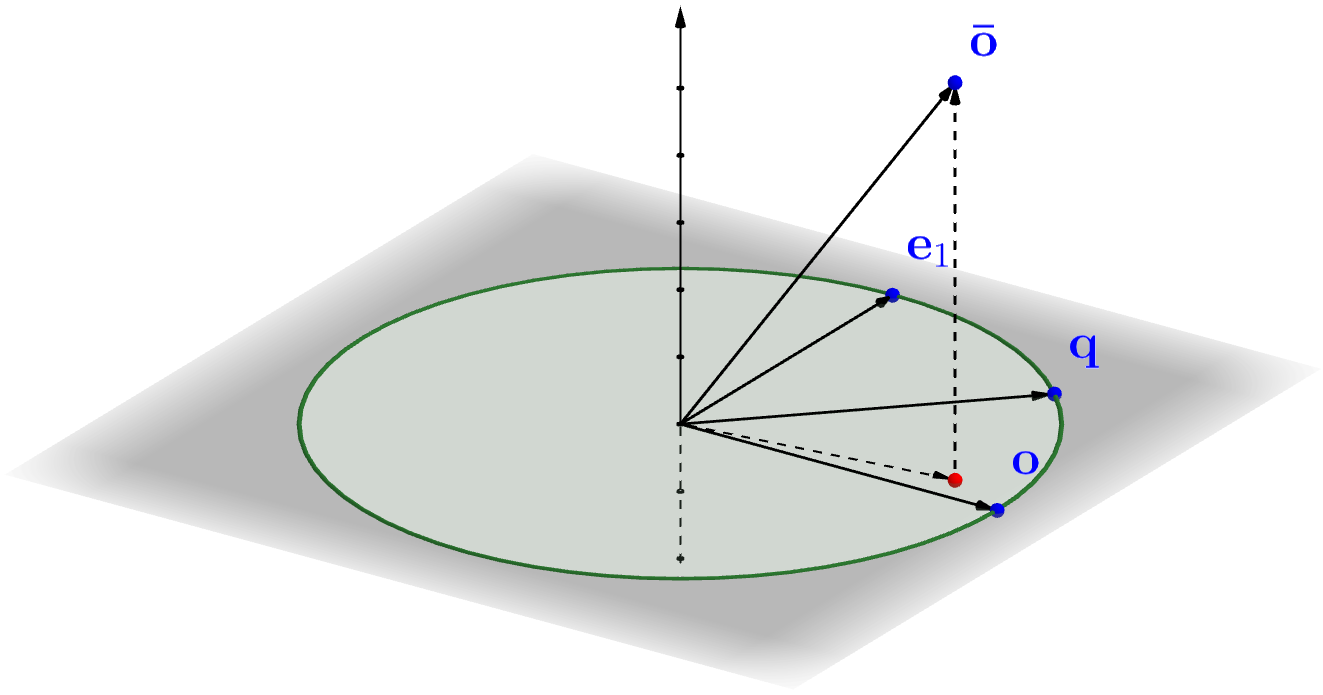}
    \end{subfigure} 
    \begin{subfigure}[b]{0.33\linewidth}
        \centering
        \includegraphics[width=\textwidth]{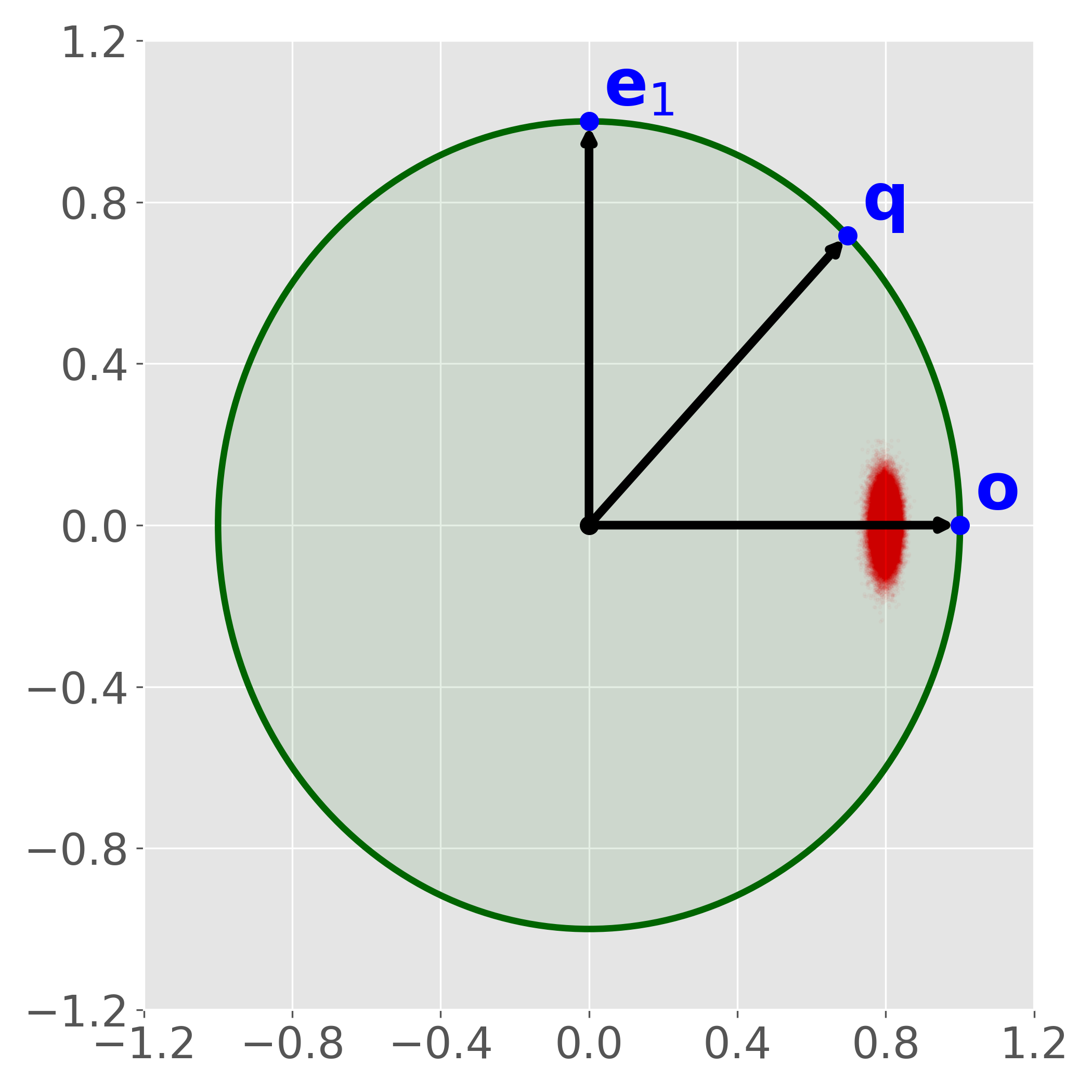}
    \end{subfigure} 
    \vspace{-2mm}
    \caption{Geometric Relationship among the Vectors.}
    \label{fig:projection}
    \vspace{-4mm}
\end{figure}
\subsubsection{Analyzing the Explicit Relationship between $\left< \mathbf{o}, \mathbf{q}  \right>$ and $\left< \mathbf{\bar o}, \mathbf{q}  \right>$}
\label{subsubsec: geometric}
We note that the relationship between $\left< \mathbf{o,q} \right> $ and $\left< \mathbf{\bar o}, \mathbf{q}\right>  $ depends only on the projection of $\mathbf{\bar o} $ on the two-dimensional subspace spanned by $\mathbf{o}$ and $\mathbf{q}$, which is illustrated on the left panel of Figure~\ref{fig:projection}. 
For the component of $\mathbf{\bar o}$ which is perpendicular to the subspace, it has no effect on the inner product $\left< \mathbf{\bar o}, \mathbf{q}\right> $.
The following lemma presents the specific result. 
The proof can be found 
in the technical report~\cite{technical_report}.
\begin{lemma}[Geometric Relationship]
    \label{lemma: geometry}
    Let $\mathbf{o}, \mathbf{q}$ and $\mathbf{\bar o}$ be 
    any three 
    unit vectors.
    When $\mathbf{o}$ and $\mathbf{q} $ are collinear (i.e., $\mathbf{o}=\mathbf{q}$ or $\mathbf{o}=-\mathbf{q} $), we have 
    \begin{align}
        \left< \mathbf{\bar o}, \mathbf{q}  \right>  = \left< \mathbf{\bar o}, \mathbf{o} \right> \cdot \left< \mathbf{o}, \mathbf{q} \right> \label{eq: collinear}
    \end{align}
    When $\mathbf{o}$ and $\mathbf{q} $ are non-collinear, we have 
    \begin{align}
        \left< \mathbf{\bar o}, \mathbf{q}  \right>  = \left< \mathbf{\bar o}, \mathbf{o} \right> \cdot \left< \mathbf{o}, \mathbf{q} \right> +   \left< \mathbf{\bar o}, \mathbf{e}_1  \right> \cdot  \sqrt {1- \left< \mathbf{o}, \mathbf{q} \right>^2 } \label{eq: non-collinear}
    \end{align}
    where $\mathbf{e}_1$ is $\mathbf{q}  - \left< \mathbf{q},\mathbf{o}  \right> \mathbf{o}$ with its norm normalized to be 1, i.e.,  $\mathbf{e}_1:= \frac{\mathbf{q}  - \left< \mathbf{q},\mathbf{o}  \right> \mathbf{o}}{\| \mathbf{q}  - \left< \mathbf{q},\mathbf{o}  \right> \mathbf{o}\|}$. We note that $\mathbf{o} \perp \mathbf{e}_1$ (since $\left< \mathbf{o}, \mathbf{e}_1\right> = 0$) and $\| \mathbf{e}_1\|=1 $.
\end{lemma}

Recall that we target to estimate $\left< \mathbf{o,q} \right> $. 
If we exactly know the values of all the variables other than $\left< \mathbf{o,q} \right> $, we can compute the exact value of $\left< \mathbf{o,q} \right> $ by solving Equations (\ref{eq: collinear}) and (\ref{eq: non-collinear}).
In particular, in Equations (\ref{eq: collinear}) and (\ref{eq: non-collinear}), $\left< \mathbf{\bar o}, \mathbf{o}  \right> $ is the inner product between the quantized data vector and the data vector. Its value can be pre-computed in the index phase.
$\left< \mathbf{\bar o}, \mathbf{q} \right> $ is the inner product between the quantized data vector and the query vector. Its value can be efficiently computed in the query phase (we will specify in Section~\ref{subsec: efficient computation} how it can be efficiently computed). 
Thus, when $\mathbf{o}$ and $\mathbf{q}$ are collinear, we can compute the value of $\left< \mathbf{o,q} \right> $ exactly by solving Equation (\ref{eq: collinear}), i.e., $\left< \mathbf{o,q} \right> = \frac{\left< \mathbf{\bar o}, \mathbf{q}  \right> }{\left< \mathbf{\bar o}, \mathbf{o}  \right> } $.

When $\mathbf{o}$ and $\mathbf{q}$ are non-collinear (which is a more common case), in order to exactly solve the Equation (\ref{eq: non-collinear}), we need to know the value of $\left< \mathbf{\bar o}, \mathbf{e}_1  \right>$. 
However, as $\mathbf{e}_1 $ depends on both $\mathbf{o}$ and $\mathbf{q}$ (which can be seen by its definition), $\left< \mathbf{\bar o}, \mathbf{e}_1  \right>$ can be neither pre-computed in the index phase (because it depends on $\mathbf{q}) $ nor computed efficiently in the query phase without accessing $\mathbf{o}$.

We notice that although we cannot efficiently compute the exact value of $\left< \mathbf{\bar o}, \mathbf{e}_1  \right> $~\footnote{In particular, when we say ``computing the value'' of a random variable, it refers to computing its \textit{observed} value based on a certain sampled $P$.}, given the random nature of $\mathbf{\bar o}$, we explicitly know its distribution. 
Specifically, recall that we have sampled a random orthogonal matrix $P$, applied it to the codebook $\mathcal{C} $ and generated a \textit{randomized} codebook $\mathcal{C}_{rand}$.
$\mathbf{\bar o}$ is a vector picked from the randomized codebook $\mathcal{C}_{rand}$ and thus, it is a random vector. $\left< \mathbf{\bar o}, \mathbf{o}  \right> $ and $ \left< \mathbf{\bar o}, \mathbf{e}_1  \right>  $ correspond to the projection of the random vector $\mathbf{\bar o} $ onto two fixed directions (i.e., the directions are $\mathbf{o} $ and $\mathbf{e}_1$, where $\mathbf{o} \perp \mathbf{e}_1 $). 
Thus, they are mutually correlated random variables.

We rigorously analyze the distributions of the random variables. 
The core conclusions of the analysis are briefly summarized as follows while the detailed presentation and proof are left in 
the technical report~\cite{technical_report}
due to the page limit.
Specifically, our analysis indicates that when $D$ ranges from $10^2$ to $10^6$, it is always true that $\left< \mathbf{\bar o}, \mathbf{o}  \right>$ has the expectation~\footnote{The exact expected value is $\mathbb{E}\left[ \left< \mathbf{\bar o}, \mathbf{o}  \right>  \right]  = \sqrt {\frac{D}{\pi}} \frac{2\Gamma (\frac{D}{2} )}{(D-1)\Gamma (\frac{D-1}{2} ) }$, where $\Gamma(\cdot)$ is the Gamma function. The expected value ranges from 0.798 to 0.800 for $D\in [10^2,10^6]$.} of around 0.8 and $\left< \mathbf{\bar o}, \mathbf{e}_1  \right> $ has the expectation of exactly 0.
It further indicates that, with high probability, these random variables would not deviate from their expectation by $\Omega(1/\sqrt {D})$. 
This conclusion quantitatively presents the extent that the random variables concentrate around their expected values, which will be used later for analyzing the error bound of our estimator.
To empirically verify our analysis, we repeatedly and independently sample the random orthogonal matrices $P$ $10^5$ times for a pair of fixed $\mathbf{o,q}$ in the 128-dimensional space. 
The right panel of Figure~\ref{fig:projection} visualizes the projection of $\mathbf{\bar o}$ on the 2-dimensional space spanned by $\mathbf{o,q}$ with the red point cloud (each point represents the projection of an $\mathbf{\bar o}$ based on a sampled random matrix $P$). 
In particular, $\left< \mathbf{\bar o}, \mathbf{o}  \right>$ (the x-axis) is shown to be concentrated around 0.8. $\left< \mathbf{\bar o}, \mathbf{e}_1  \right> $ (the y-axis) is concentrated and symmetrically distributed around 0, which verifies our theoretical analysis perfectly. 

\subsubsection{Constructing an Unbiased Estimator for $\left< \mathbf{o,q} \right> $}
\label{subsubsec: construct unbiased estimator}
Based on our analysis on Equation (\ref{eq: collinear}), for the case that $\mathbf{o,q}$ are collinear, $\left< \mathbf{o,q} \right> $ can be explicitly solved by $\left< \mathbf{o,q} \right> = \frac{\left< \mathbf{\bar o}, \mathbf{q}  \right> }{\left< \mathbf{\bar o}, \mathbf{o}  \right> } $.
Thus, it is natural to conjecture that for the case that $\mathbf{o,q}$ are non-collinear, $\frac{\left< \mathbf{\bar o}, \mathbf{q}  \right> }{\left< \mathbf{\bar o}, \mathbf{o}  \right> }$ should also be a good estimator for $\left< \mathbf{o,q} \right> $. We thus deduce from it as follows.
\begin{align}
    \frac{\left< \mathbf{\bar o}, \mathbf{q}  \right> }{\left< \mathbf{\bar o}, \mathbf{o}  \right> } 
        = & \frac{\left< \mathbf{\bar o}, \mathbf{o} \right> \cdot \left< \mathbf{o}, \mathbf{q} \right> +   \left< \mathbf{\bar o}, \mathbf{e}_1  \right> \cdot  \sqrt {1- \left< \mathbf{o}, \mathbf{q} \right>^2 }}{\left< \mathbf{\bar o}, \mathbf{o}  \right> } \label{eq: plugin geometry} \\
        = &\left< \mathbf{o, q} \right> + \sqrt {1 - \left< \mathbf{o,q} \right>^2 } \cdot  \frac{\left< \mathbf{\bar o}, \mathbf{e}_1  \right> }{ \left< \mathbf{\bar o}, \mathbf{o}  \right> } \label{eq: simplify}
\end{align}
where Equation (\ref{eq: plugin geometry}) is by Equation (\ref{eq: non-collinear}) and
Equation (\ref{eq: simplify}) simplifies Equation (\ref{eq: plugin geometry}). 
We note that the last term in Equation (\ref{eq: simplify}) can be viewed as the error term of the estimator. Recall that based on our analysis in Section~\ref{subsubsec: geometric}, $\left< \mathbf{\bar o}, \mathbf{o}  \right> $ is concentrated around 0.8. $\left< \mathbf{\bar o}, \mathbf{e}_1  \right> $ has the expectation of 0 and is concentrated. It implies that the error term has 0 expectation and will not deviate largely from 0 due to the concentration. The following theorem presents the specific results. The rigorous proof can be found in 
the technical report~\cite{technical_report}.

\begin{theorem}[Estimator]
    \label{theorem: estimator}
    The unbiasedness is given as 
    \begin{align}
        \mathbb{E} \left[ \frac{\left< \mathbf{\bar o}, \mathbf{q}  \right> }{\left< \mathbf{\bar o}, \mathbf{o}  \right> }  \right]  = \left< \mathbf{o}, \mathbf{q}  \right> \label{eq: unbiasedness}
    \end{align} 
    The error bound of the estimator is given as
    \begin{align}
        \mathbb{P} \left\{ \left| \frac{\left< \mathbf{\bar o}, \mathbf{q}  \right> }{\left< \mathbf{\bar o,\mathbf{o} } \right> } -\left< \mathbf{o,q} \right>   \right| >  \sqrt{\frac{{1 - \left< \mathbf{\bar o}, \mathbf{o}  \right>^2}}{\left< \mathbf{\bar o}, \mathbf{o}  \right>^2 }}\cdot \frac{\epsilon_0}{\sqrt {D-1} }   \right\}  \le 2e^ { -c_0 \epsilon_0^2} \label{eq:error bound}
    \end{align}
    where $\epsilon_0$ is a parameter which controls the failure probability. $c_0$ is a constant factor. 
    The error bound can be concisely presented as
    \begin{align}
        \left| \frac{\left< \mathbf{\bar o}, \mathbf{q}  \right> }{\left< \mathbf{\bar o,\mathbf{o} } \right> } -\left< \mathbf{o,q} \right>   \right| = O \left( \frac{1}{\sqrt {D}} \right) \ with\ high\ probability
        \label{eq: concise presentation of error bound}
    \end{align}
\end{theorem}

Due to Equations (\ref{eq:dis to ip}) and (\ref{eq: unbiasedness}), the unbiased estimator of $\left< \mathbf{o,q} \right> $ can further induce an unbiased estimator of the squared distance between the raw data and query vectors. We provide empirical verification on the unbiasedness in Section~\ref{subsubsec: verify unbiasedness}. 
Besides, we would like to highlight that based on similar analysis, an alternative estimator
$\left< \mathbf{o}, \mathbf{q}  \right> \approx \left< \mathbf{\bar o}, \mathbf{q}  \right> $, i.e., by simply treating the quantized data vector as the data vector as PQ does, can be easily proved to be \textit{biased}.

Equation (\ref{eq:error bound}) presents the error bound of our estimator. 
In particular, it presents a $1-2\exp (-c_0 \epsilon_0^2)$ confidence interval
\begin{align}
    \frac{\left< \mathbf{\bar o}, \mathbf{q}  \right> }{\left< \mathbf{\bar o,\mathbf{o} } \right> }  \pm \sqrt{\frac{{1 - \left< \mathbf{\bar o}, \mathbf{o}  \right>^2}}{\left< \mathbf{\bar o}, \mathbf{o}  \right>^2 }}\cdot \frac{\epsilon_0}{\sqrt {D-1} }
\end{align}
We note that the failure probability (i.e., the probability that the confidence interval does not cover the true value of $\left< \mathbf{o,q} \right> $) is $2\exp (-c_0 \epsilon_0^2)$. It decays in a quadratic-exponential trend wrt $\epsilon_0$, which is extremely fast. The length of the confidence interval grows linearly wrt $\epsilon_0$. 
Thus, $\epsilon_0=\Theta(\sqrt {\log (1/\delta)})$ corresponds to a failure probability of at most $\delta$, which indicates that a short confidence interval can correspond to a high confidence level. 
In practice, $\epsilon_0$ is fixed to be 1.9 in pursuit of nearly perfect confidence (see Section~\ref{subsubsec: verify epsilon0} for the empirical verification study).
Recall that $\left< \mathbf{\bar o},\mathbf{o}  \right> $ is concentrated around 0.8. 
Based on the values of $\epsilon_0$ and $\left< \mathbf{\bar o}, \mathbf{o}  \right> $, the error bound can be further concisely presented as Equation (\ref{eq: concise presentation of error bound}), i.e., it guarantees an error bound of $O(1/\sqrt {D})$. According to a recent theoretical study~\cite{2017_focs_additive_error}, for $D$-dimensional vectors, with a short code of $D$ bits, it is \textit{impossible} in theory for a method to provide a bound which is tighter than $O(1/\sqrt{D})$ (the failure probability is viewed as a constant). Thus, Equation (\ref{eq: concise presentation of error bound}) indicates that RaBitQ's error bound is sharp, i.e., it achieves the asymptotic optimality.
The error bound will be later used in ANN search to determine whether a data vector should be re-ranked (see Section~\ref{sec: RaBitQ for ANN}). 

Furthermore, we note that RaBitQ provides an error bound in an additive form~\cite{2017_focs_additive_error} (i.e., absolute error). When the data vectors are well normalized (recall that in Section~\ref{subsubsec: pre-procession} we normalize the data vectors), the bound can be pushed forward to a multiplicative form~\cite{SIAM_JOC_2022_indyk} (i.e., relative error).
We leave the detailed discussion in 
the technical report~\cite{technical_report}
because it is based on an assumption that the data vectors are well normalized. 
Note that all other theoretical results introduced in this paper do not rely on any assumptions on the data, i.e., the additive bound holds regardless of the data distribution.
In the present work, we adopt a simple and natural method of normalization (i.e., with the centroids of IVF as will be introduced in Section~\ref{sec: RaBitQ for ANN}) to instantiate our scheme of quantization, while we have yet to extensively explore the normalization step itself. We shall leave it as future work to rigorously study the normalization problem.

\subsection{Computing the Estimator Efficiently}
\label{subsec: efficient computation}
Recall that $\frac{\left< \mathbf{\bar o}, \mathbf{q}  \right> }{\left< \mathbf{\bar o}, \mathbf{o}  \right> }$ is the estimator. Since $\left< \mathbf{\bar o}, \mathbf{o}  \right>$ has been pre-computed during the index phase, 
the remaining question is to compute the value of $\left< \mathbf{\bar o}, \mathbf{q}  \right> $ efficiently. 
For the sake of convenience, we denote $P^{-1}\mathbf{q}$ as $\mathbf{q}'$.
Like what we do in Section~\ref{subsubsec: quantization code}, in order to compute $\left< \mathbf{\bar o}, \mathbf{q}  \right> $, 
we can compute $\left< \mathbf{\bar x}, \mathbf{q}' \right> $, which can be verified as follows.
\begin{align}
    \left< \mathbf{\bar o}, \mathbf{q}  \right> =\left< P \mathbf{\bar x}, \mathbf{q}  \right> =\left< P^{-1}P\mathbf{\bar x}, P^{-1} \mathbf{q}  \right> = \left< \mathbf{\bar x}, \mathbf{q}'   \right>
\end{align}

\subsubsection{Quantizing the Transformed Query Vector}
\label{subsubsec: quantize query vector}
Recall that $\mathbf{\bar x}$ is a bi-valued vector whose entries are $\pm 1/\sqrt {D} $. It is represented and stored as a binary quantization code 
$\mathbf{\bar x}_b$ as is discussed in Section~\ref{subsubsec: quantization code}. 
$\mathbf{q}'$ is a real-valued vector, whose entries are conventionally represented by floating-point numbers (floats in short). 
We note that in our method, representing the entries of $\mathbf{q}'$ with floats is an overkill. 
Specifically, recall that our method adopts $\frac{\left< \mathbf{\bar o}, \mathbf{q} \right> }{\left< \mathbf{\bar o}, \mathbf{o}  \right> } $ as an estimator of $\left< \mathbf{o,q}\right> $. 
Even if we obtain a perfectly accurate result in the computation of $\left< \mathbf{\bar o}, \mathbf{q}  \right> $, our estimation of $\left< \mathbf{o,q} \right> $ is still approximate.
Thus, instead of exactly computing $\left< \mathbf{\bar o}, \mathbf{q} \right> $, we aim to guarantee that the error produced in the computation of $\left< \mathbf{\bar o}, \mathbf{q}  \right> $ is much smaller than the error of the estimator itself. 

Specifically, we apply uniform scalar quantization on the entries of $\mathbf{q}'$ and represent them as $B_q$-bit unsigned integers.
We denote the $i$th entry of the vector $\mathbf{q}'$ as $\mathbf{q}'[i]$. 
Let $v_l:= \min_{1\le i\le D} \mathbf{q}'[i], v_r:= \max_{1\le i\le D} \mathbf{q}'[i]$ and $\Delta := (v_r - v_l)/(2^{B_q}-1)$.
The uniform scalar quantization uniformly splits the range of the values $[v_l, v_r]$ into $2^{B_q}-1$ segments, where each segment has the length of $\Delta$.
Then for a value $v=v_l + m\cdot \Delta + t, m=0,1,...,2^{B_q}-1, t \in [0, \Delta)$, the method quantizes it by rounding it up to its nearest boundary of the segments (i.e., 
$v_l + m \cdot \Delta$ or $v_l + (m+1) \cdot \Delta$)  and representing it with the corresponding $B_q$-bit unsigned integer (i.e., $m$ or $m+1$). 
Let $\mathbf{\bar q}$ be the vector whose entries are equal to the quantized values of the entries of $\mathbf{q}'$ (we term it as the quantized query vector) and $\mathbf{\bar q}_u$ be its $B_q$-bit unsigned integer representation, where 
$\mathbf{\bar q}= \Delta \cdot \mathbf{\bar q}_u + v_l \cdot \mathbf{1}_D$ (recall that $\mathbf{1}_D$ is the $D$-dimensional vector with all its entries as ones).
Then, we can compute $\left< \mathbf{\bar x}, \mathbf{\bar q}  \right> $ as an approximation of $\left<\mathbf{\bar x}, \mathbf{q'}  \right> $.

Furthermore, to retain the theoretical guarantee, we adopt the trick of randomizing the uniform scalar quantization~\cite{randomized_uniform_quantization, 2017_focs_additive_error}. 
Specifically, unlike the conventional method which rounds up a value to its nearest boundary of the segments, the randomized method rounds it to its left or right boundary randomly.
The rationale is that for a value $v=v_l + m\cdot \Delta + t, m=0,1,...,2^{B_q}-1, t \in [0, \Delta)$, when it is rounded to $v_l + m\cdot \Delta$, it will cause an error of under-estimation $-t<0$. When it is rounded to $v_l + (m+1)\cdot \Delta$, it will cause an error of over-estimation $\Delta-t>0$. 
If we assign $1-t/\Delta$ probability to the former event and $t/\Delta$ probability to the latter event, the expected error would be 0, which makes the computation unbiased.
We note that this operation can be easily achieved by letting
\begin{align}
    \mathbf{\bar q}_u[i] := \Big\lfloor \frac{\mathbf{q}'[i]-v_l }{\Delta} + u_i  \Big\rfloor  
\end{align}
where $u_i$ is sampled from the uniform distribution on $[0,1]$. 
Moreover, based on the randomized method, we can analyze the minimum $B_q$ needed for making the error introduced by the uniform scalar quantization negligible.
The result is presented with the following theorem.
The detailed proof can be found in 
the technical report~\cite{technical_report}.
\begin{theorem}
    $B_q=\Theta (\log \log D)$ suffices to guarantee that $|\left< \mathbf{\bar x},\mathbf{q}'  \right> - \left< \mathbf{\bar x},\mathbf{\bar q}  \right>| = O(1/\sqrt {D})$ with high probability.
    \label{theorem: 3.3}
\end{theorem}
Recall that the estimator has the error of $O(1/\sqrt {D})$ (see Section~\ref{subsubsec: construct unbiased estimator}).
The above theorem shows that setting $B_q=\Theta (\log \log D)$ suffices to guarantee that the error introduced by the uniform scalar quantization is at the same order as the error introduced by estimator. 
Because the error decreases exponentially wrt $B_q$, increasing $B_q$ by a small constant (i.e., $B_q$ is still at the order of $\Theta (\log \log D)$) guarantees that the error is much smaller than that of the estimator.
We provide the empirical verification study for $B_q$ in Section~\ref{subsubsec: verify BQ}. The result shows that when $B_q=4$, the error introduced by the uniform scalar quantization would be negligible.

\subsubsection{Computing $\left< \mathbf{\bar x}, \mathbf{\bar q}  \right> $ Efficiently}
\label{subsubsec: efficient computation of quantized inner product}
We next present how to compute $\left< \mathbf{\bar x}, \mathbf{\bar q}  \right> $ efficiently. We first express $\left< \mathbf{\bar x}, \mathbf{\bar q}  \right> $ with $\mathbf{\bar x}_b, \mathbf{\bar q}_u$ as follows.
\begin{align}
    &\left< \mathbf{\bar x}, \mathbf{\bar q}  \right> = \left< \frac{2 \mathbf{\bar x}_b-\mathbf{1}_D }{\sqrt {D}} , \Delta \cdot \mathbf{\bar q}_u + v_l  \cdot \mathbf{1}_D \right> \\
    =  &\frac{2\Delta}{\sqrt {D} } \left< \mathbf{\bar x}_b, \mathbf{\bar q}_u  \right> + \frac{2v_l}{\sqrt {D}} \sum_{i=1}^{D} \mathbf{\bar x}_b[i]   
    - \frac{\Delta}{\sqrt {D} }  \sum_{i=1}^{D} \mathbf{\bar q}_u[i] - \sqrt {D} \cdot v_l \label{eq:compute inner product}
\end{align}
Note that the factors $\Delta$ and $v_l$ are known when we quantize the query vector. 
$\sum_{i=1}^{D} \mathbf{\bar x}_b[i] $ corresponds to the number of $1$'s in 
the bit string $\mathbf{\bar x}_b$, which can be pre-computed during the index phase. 
$\sum_{i=1}^{D} \mathbf{\bar q}_u[i]$ depends only on the query vector. Its cost of computation can be shared by all the data vectors. The remaining task is to compute $ \left< \mathbf{\bar x}_b, \mathbf{\bar q}_u  \right>$ where the coordinates of $\mathbf{\bar x}_b$ are $0$ or $1$ and those of $\mathbf{\bar q}_u $ are unsigned $B_q$-bit integers. 

We provide two versions of fast computation for $ \left< \mathbf{\bar x}_b, \mathbf{\bar q}_u  \right>$. 
The first version targets the case of a \emph{single} quantization code, as the original implementation of PQ~\cite{jegou2010product} does. 
The second version targets the case of a packed \emph{batch} of quantization codes, as the fast SIMD-based implementation of PQ~\cite{fastscan,fastscanavx2} does.
We note that in general, both our method and PQ have higher throughput in the second case than that in the first case, i.e., they estimate the distances for more quantization codes within certain time.
We note that the second case requires the quantization codes to be packed in a batch, which is feasible in some certain scenarios only.

\begin{figure}[thb]
    \vspace{-2mm}
    \centering
    \includegraphics[width=0.9\linewidth]{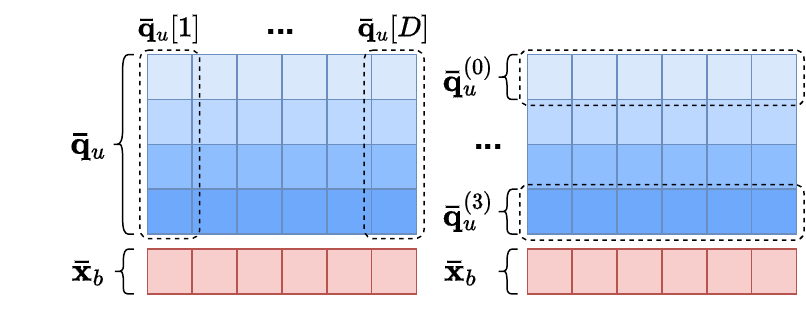}
    \vspace{-2mm}
    \caption{Bitwise Decomposition of $\mathbf{\bar q}_u$.}
    \label{fig:bitwise decomposition}
    \vspace{-2mm}
\end{figure}

For the first case where the estimation of the distance is for a query vector and a \textit{single} quantization code, we note that an unsigned $B_q$-bit integer can be decomposed into $B_q$ binary values as shown in Figure~\ref{fig:bitwise decomposition}. 
The left panel represents the naive computation of $ \left< \mathbf{\bar x}_b, \mathbf{\bar q}_u  \right>$. 
The right panel represents the proposed bitwise computation of $ \left< \mathbf{\bar x}_b, \mathbf{\bar q}_u  \right>$.
Let $\mathbf{\bar q}_{u}^{(j)}[i] \in \{0,1\}$ be the $j$th bit of $\mathbf{\bar q}_{u}[i]$ where $0\le j < B_q$. The following expression specifies the idea. 
\begin{align}
    \left< \mathbf{\bar x}_b, \mathbf{\bar q}_u  \right> = &\sum_{i=1}^{D} \mathbf{\bar x}_b[i] \cdot  \mathbf{\bar q}_u[i] 
    = \sum_{i=1}^{D}  \mathbf{\bar x}_b[i] \cdot  \sum_{j=0}^{B_q-1}\mathbf{\bar q}_{u}^{(j)}[i] \cdot 2^j \\
    = &\sum_{j=0}^{B_q-1}2^j\cdot \sum_{i=1}^{D}  \mathbf{\bar x}_b[i] \cdot  \mathbf{\bar q}_{u}^{(j)}[i] = \sum_{j=0}^{B_q-1}2^j\cdot \left<\mathbf{\bar x}_b, \mathbf{\bar q}_{u}^{(j)}\right> \label{eq: efficient computation}
\end{align}
Equation (\ref{eq: efficient computation}) shows that $\left< \mathbf{\bar x}_b, \mathbf{\bar q}_u  \right>$ can be expressed as a weighted sum of the inner product of the binary vectors, i.e., $\left<\mathbf{\bar x}_b, \mathbf{\bar q}_{u}^{(j)}\right>$ for $0\le j < B_q$. 
In particular, we note that the inner product of binary vectors can be efficiently achieved by bitwise operations, i.e., bitwise-and and popcount (a.k.a., bitcount).
Thus, the computation of $\left< \mathbf{\bar o}, \mathbf{q}  \right> $
is finally reduced to $B_q$ bitwise-and and popcount operations on $D$-bit strings, which are well supported by virtually all platforms.
As a comparison, we note that, as is comprehensively studied in \cite{fastscan}, PQ relies on looking up LUTs in RAM, which cannot be implemented efficiently. Based on our experiments in Section~\ref{subsubsec: time-accuracy trade-off per vector}, on average our method runs 3x faster than PQ and OPQ (a popular variant of PQ~\cite{ge2013optimized}) while reaching the same accuracy. 

For the second case where the estimation of the distance is for a query vector and a packed \emph{batch} of quantization codes, we note that our method can seamlessly adopt the same fast SIMD-based implementation~\cite{fastscan,fastscanavx2} as PQ does. 
In particular, for a $D$-bit string, we split it into $D/4$ sub-segments where each sub-segment has 4 bits. 
We then pre-process $D/4$ LUTs where each LUT has $2^4$ unsigned 
integers corresponding to the inner products between a sub-segment of $\mathbf{\bar q}_u$ and the $2^4$ possible binary strings of a $4$-bit sub-segment.
For a quantization code of a data vector, we can compute $ \left< \mathbf{\bar x}_b, \mathbf{\bar q}_u  \right>$ by looking up and accumulating the values in the LUTs for $D/4$ times.
We note that the computation is reduced to exactly the form of PQ and thus can adopt the fast SIMD-based implementation seamlessly. 
Recall that our method has the quantization codes of $D$ bits by default while PQ and OPQ have the codes of $2D$ bits by default. Therefore, our method has better efficiency than PQ and OPQ for computing approximate distances based on quantized vectors.
Furthermore, as is shown in Section~\ref{subsubsec: time-accuracy trade-off per vector}, in the default setting, our method also achieves consistently better accuracy than PQ and OPQ despite that our method uses a shorter quantization code (i.e., $D$ v.s. $2D$).

\subsection{Summary of RaBitQ}
\label{subsec: summary}

We summarize the RaBitQ algorithm in Algorithm~\ref{code: RaBitQ index phase} (the index phase) and Algorithm~\ref{code: RaBitQ query phase} (the query phase). 
In the index phase, it takes a set of raw data vectors as inputs. It normalizes the set of vectors based on Section~\ref{subsubsec: pre-procession} (line 1), constructs the RaBitQ codebook by sampling a random orthogonal matrix $P$ based on Section~\ref{subsubsec: the RaBitQ} (line 2) and computes the quantization codes $\mathbf{\bar x}_b$ based on Section~\ref{subsubsec: quantization code} (line 3).
In the query phase, it takes a raw query vector, a set of IDs of the data vectors and the pre-processed variables about the RaBitQ method as inputs. It first inversely transforms, normalizes and quantizes the raw query vector (line 1-2). We note that the time cost of these steps can be shared by all the data vectors. Then for each input ID of the data vectors, it efficiently computes the value of $\frac{\left< \mathbf{\bar o}, \mathbf{q}  \right> }{\left< \mathbf{\bar o}, \mathbf{o}  \right>} $ based on Section~\ref{subsubsec: efficient computation of quantized inner product}, adopts it as an unbiased estimation of $\left< \mathbf{o,q} \right> $ based on Section~\ref{subsec: unbiased estimator} and further computes an estimated distance between the raw query and the raw data vectors based on Section~\ref{subsubsec: pre-procession} (line 3-5).

\begin{algorithm}[tbh]
\DontPrintSemicolon
\SetKwData{and}{and}
\SetKwInOut{Input}{Input}\SetKwInOut{Output}{Output}
\Input{A set of raw data vectors}
\Output{
The sampled matrix $P$; 
the quantization code $\mathbf{\bar x}_b $; the pre-computed results of $\| \mathbf{o}_r-\mathbf{c} \|$ and $\left< \mathbf{\bar o}, \mathbf{o}  \right>$ 
}
\BlankLine
Normalize the set of vectors (Section~\ref{subsubsec: pre-procession})\;
Sample a random orthogonal matrix $P$ to construct the codebook $\mathcal{C}_{rand}$ (Section~\ref{subsubsec: the RaBitQ})\;
Compute the quantization codes $\mathbf{\bar x}_b$ (Section~\ref{subsubsec: quantization code})\;
Pre-compute the values of $\| \mathbf{o}_r-\mathbf{c} \|$ and $\left< \mathbf{\bar o}, \mathbf{o}  \right>$ \;
\caption{RaBitQ (Index Phase)}
\label{code: RaBitQ index phase}
\end{algorithm}
\begin{algorithm}[tbh]
\DontPrintSemicolon
\SetKwData{and}{and}
\SetKwInOut{Input}{Input}\SetKwInOut{Output}{Output}
\Input{A raw query vector $\mathbf{q}_r$; the sampled matrix $P$; 
a set of IDs of the data vectors, their quantization codes $\mathbf{\bar x}_b$ and the results of $\| \mathbf{o}_r-\mathbf{c} \| $ and $ \left< \mathbf{\bar o}, \mathbf{o}  \right>$}
\Output{A set of approximate distances between the raw query and the raw data vectors}
\BlankLine
Normalize and inversely transform the raw query vector and obtain $\mathbf{q}'$\;
Quantize $\mathbf{q}'$ into $\mathbf{\bar q}$ (Section~\ref{subsubsec: quantize query vector})\;
\ForEach{input ID of the data vectors}{%
    Compute the value of the estimator $\frac{\left< \mathbf{\bar o}, \mathbf{q}  \right>}{\left< \mathbf{\bar o}, \mathbf{o}  \right>} $
    as an approximation of $\left< \mathbf{o,q} \right> $ 
    (Section~\ref{subsubsec: efficient computation of quantized inner product})\;
    Compute an estimated distance between the raw query and the data vector based on Equation~(\ref{eq:dis to ip})\;
}
\caption{RaBitQ (Query Phase)}
\label{code: RaBitQ query phase}
\end{algorithm}

\section{RaBitQ for In-Memory ANN Search}
\label{sec: RaBitQ for ANN}

Next we present the application of our method to the in-memory ANN search. 
We note that the popular quantization method PQx4fs has been used in combination with the inverted-file-based indexes such as IVF~\cite{jegou2010product} or the graph-based indexes such as NGT-QG~\cite{NGT-QG} for in-memory ANN search.
The combination of a quantization method with IVF can be easily done without much efforts. For example, we can use the quantization method to estimate the distances between the data vectors in the clusters that are probed, which decide those vectors to be re-ranked based on exact distances. In this case, batches of data vectors can be formed and the SIMD-based fast implementation (i.e., PQx4fs) can be adopted. Nevertheless, the combination of a quantization method with graph-based methods such as NGT-QG would require much more efforts in order to make the combined method work competitively in the in-memory setting, which would be of independent interest. This is because in graph-based methods, the vectors to be searched are decided one after one based on the greedy search process in the run-time, and it is not easy to form batches of them so that SIMD-based fast implementation can be adopted. 
Therefore, we apply our method in combination with IVF index~\cite{jegou2010product} in this paper. 
We leave it as future work to apply our quantization method in graph-based methods.

We present the workflow of the RaBitQ method with the IVF index as follows.
During the \underline{index phase}, for a set of raw data vectors, the IVF algorithm first clusters them with the KMeans algorithm, builds a bucket for each cluster and assigns the vectors to their corresponding buckets. 
Our method then normalizes the raw data vectors based on the centroids of their corresponding clusters and feeds the set of the normalized vectors to the subsequent steps of our RaBitQ method. 
During the \underline{query phase}, for a raw query vector, the algorithm selects the first $N_{probe}$ clusters whose centroids are nearest to the query. Then for each selected cluster, the algorithm retrieves all the quantization codes and estimates their distances based on the quantization codes, which decide the vectors to be re-ranked based on exact distances.

As for re-ranking~\cite{learningtohashsurvey}, PQ and its variants set a fixed hyper-parameter which decides the number of data vectors to be re-ranked (i.e., they re-rank the vectors with the smallest estimated distances).
Specifically, they retrieve their raw data vectors, compute the exact distances and find out the final NN.
In particular, the tuning of the hyper-parameter is empirical and often hard as it can vary largely across different datasets (see Section~\ref{subsubsec: time-accuracy trade-off ann search}).
In contrast, recall that in our method, there is a sharp error bound as discussed in Section~\ref{subsec: unbiased estimator} (note that the error bound is rigorous and always holds regardless of the data distribution). 
Thus, we decide the data vectors to be re-ranked based on the error bound without tuning hyper-parameters. 
Specifically, if a data vector has its lower bound of the distance greater than the exact distance of the currently searched nearest neighbor, then we drop it.
Otherwise, we compute its exact distance for re-ranking. 
Due to the theoretical error bound, the re-ranking strategy has the guarantee of correctly sending the true NN from the probed clusters to re-ranking with high probability.
The empirical verification can be found in Section~\ref{subsubsec: verify epsilon0}.
We emphasize that the idea of re-ranking based on a bound is not new. There are many studies from the database community that adopt a similar strategy~\cite{revision1_bound, revision2_bound, dataseries_benchmark, vafile, vaplusfile, wang2023dumpy} for improving the robustness of similarity search for various data types. 
We note that beyond the idea of re-ranking based on an error bound, RaBitQ provides rigorous theoretical 
analysis on the tightness of the bounds 
and achieves the asymptotic optimality as we have discussed in Section~\ref{subsubsec: construct unbiased estimator}.

Moreover, it is worth noting that re-ranking is a vital step for pushing forward RaBitQ's rigorous error bounds on the distances to the robustness of ANN search. In particular, when the ANN search requires higher accuracy than what RaBitQ can guarantee (e.g., when the true distances from the query to two different data vectors are extremely close to each other), then the estimated distance produced by RaBitQ would be less effective to rank them correctly. 
Re-ranking, in this case, is necessary for achieving high recall. 
Note that it is inherently difficult for any methods of distance estimation when the distances are extremely close to each other. 

\section{experiments}
\label{sec:experiments}

\subsection{Experimental Setup}
\label{subsec: experimental setup}
Our experiments involve three folds. \underline{First}, 
we compare our method with the conventional quantization methods in terms of the time-accuracy trade-off of distance estimation and time cost of index phase (with results shown in Section~\ref{subsubsec: time-accuracy trade-off per vector} and Section~\ref{subsubsec: indexing time}). 
\underline{Second}, we compare the methods when applied for in-memory ANN (with results shown in Section~\ref{subsubsec: time-accuracy trade-off ann search}).
For ANN, we target to retrieve the 100 nearest neighbors for each query, i.e., $K=100$, by following~\cite{fu2019fast}. 
\underline{Third}, we empirically verify our theoretical analysis (with results shown in Section~\ref{subsubsec: verify epsilon0} to \ref{subsubsec: verify unbiasedness}). 
\underline{Finally}, we note that RaBitQ is a method with rigorous theoretical guarantee. Its components are an integral whole and together explain its asymptotically optimal performance. The ablation of any component would cause the loss of the theoretical guarantee (i.e., the method becomes heuristic and the performance is no more theoretically predictable) and further disables the error-bound-based re-ranking (Section~\ref{sec: RaBitQ for ANN}). 
Despite this, we include empirical ablation studies in 
the technical report~\cite{technical_report}.

\smallskip\noindent\textbf{Datasets.} We use six public real-world datasets with varying sizes and dimensionalities, whose details can be found in Table~\ref{tab:data}. These datasets have been widely used to benchmark ANN algorithms~\cite{annbenchmark,lu2021hvs,lsq++,li2019approximate}.
In particular, it has been reported that on the datasets SIFT, DEEP and GIST, PQx4fs and OPQx4fs have good empirical performance~\cite{fastscanavx2}.
We note that all these public datasets provide both data and query vectors.

\begin{table}[h]
\vspace{-3mm}
\caption{Dataset Statistics}
\vspace{-4mm}
\label{tab:data}
\begin{tabular}{c|cccc}
\hline
Dataset & Size      & $D$ &  Query Size & Data Type \\ \hline
Msong   & 992,272   & 420       &  200     & Audio     \\
SIFT  & 1,000,000 & 128       &  10,000     & Image      \\
DEEP    & 1,000,000 & 256       &  1,000     & Image     \\
Word2Vec & 1,000,000 & 300      &  1,000      & Text      \\
GIST    & 1,000,000 & 960       &  1,000     & Image     \\
Image  & 2,340,373 & 150       &  200     & Image     \\ \hline
\end{tabular}
\vspace{-3mm}
\end{table}

\smallskip\noindent\textbf{Algorithms.} 
\underline{First}, 
for estimating the distances between data vectors and query vectors, we consider three baselines, PQ~\cite{jegou2010product}, OPQ~\cite{ge2013optimized} and LSQ~\cite{lsq,lsq++}.
In particular, \textbf{(1) PQ} and \textbf{(2) OPQ} are the most popular methods among the quantization methods~\cite{jegou2010product,ge2013optimized}.
They are widely deployed in industry~\cite{johnson2019billion_faiss, PASE, milvus,apple}. The popularity of PQ and OPQ indicates that they have been
empirically evaluated to the widest extent and are expected to have the best known stability.
Thus, we adopt PQ and OPQ as the primary baseline methods representing the quantization methods which have no theoretical error bounds.
There is another line of the quantization methods named the additive quantization~\cite{additivePQ,composite_quantization,lsq,lsq++}. 
Compared with PQ, these methods target extreme accuracy at the cost of much higher time for optimizing the codebook and mapping the data vectors into quantization codes in the index phase. 
\textbf{(3) LSQ}~\cite{lsq,lsq++} is the state-of-the-art method of this line.
Thus, we adopt LSQ as the baseline method representing the quantization methods which pursue extreme performance in the query phase.
The baseline methods are taken from the 1.7.4 release version of the open-source library Faiss~\cite{johnson2019billion_faiss}, which is well-optimized with the SIMD instructions of AVX2. 
\underline{Second}, for ANN, we compare our method with the most competitive baseline method OPQ according to the results in Section~\ref{subsubsec: time-accuracy trade-off per vector}.
For both our method and OPQ, we combine them with the IVF index as specified in Section~\ref{sec: RaBitQ for ANN}.
We also include the comparison with \textbf{HNSW}~\cite{malkov2018efficient} as a reference. It is one of the state-of-the-art graph-based methods as is benchmarked in \cite{graphbenchmark, annbenchmark} and is also widely adopted in industry~\cite{johnson2019billion_faiss,milvus,PASE,apple}. The implementation is taken from the hnswlib~\cite{malkov2018efficient} optimized with the SIMD instructions of AVX2.
We note that a recent quantization method ScaNN~\cite{scann} proposes a new objective function for constructing the quantization codebook of PQ and claims better empirical performance. However, as has been reported~\footnote{\url{https://github.com/facebookresearch/faiss/wiki/Indexing-1M-vectors}}, its superior performance is mainly due to the fast SIMD-based implementation~\cite{fastscanavx2}. The advantage vanishes when PQ is implemented with the same technique. Thus, we exclude it from the comparison.
Furthermore, we exclude the comparison with the LSH methods because it has been reported that the quantization methods outperform these methods empirically by orders of magnitudes in efficiency when reaching the same recall~\cite{li2019approximate}. 
The latest advances in LSH have not changed this trend~\cite{dblsh}.
Thus, comparable performance with the quantization methods indicates significant improvement over the LSH methods. 

\smallskip\noindent\textbf{Performance Metrics.}
\underline{First}, for estimating the distances between data vectors and query vectors, we use two metrics to measure the accuracy and one metric to measure the efficiency.
In particular, we measure the accuracy with (1) the average relative error and (2) the maximum relative error on the estimated squared distances. The former measures the general quality of the estimated distances while the latter measures the robustness of the estimated distances. We measure the efficiency with the time for distance estimation per vector. We note that due to the effects of cache, the efficiency depends on the order of estimating distances for the vectors. To simulate the order when the methods are used in practice, we build the IVF index for all methods and estimate the distances in the order that the IVF index probes the clusters. We measure the end-to-end time of estimating distances for all the quantization codes in a dataset and divide it by the size of the dataset. 
We take the pre-processing time in the query phase (e.g., the time for normalizing, transforming and quantizing the query vector for our method) into account, thus making the comparisons fair.
We also measure the time costs of the methods in the index phase. 
\underline{Second}, for ANN, we adopt recall and average distance ratio for measuring the accuracy of ANN search. Specifically, recall is the ratio between the number of retrieved true nearest neighbors over $K$. Average distance ratio is the average of the distance ratios of the returned $K$ data vectors wrt the ground truth nearest neighbors. 
These metrics are widely adopted to measure the accuracy of ANN algorithms~\cite{li2019approximate,annbenchmark,c2lsh,sun2014srs,huang2015query}. 
We adopt query per second (QPS), i.e., the number of queries a method can handle in a second, to measure the efficiency. It is widely adopted to measure the efficiency of ANN algorithms~\cite{li2019approximate, graphbenchmark, annbenchmark}.
Following \cite{li2019approximate, graphbenchmark, annbenchmark}, the query time is evaluated in a single thread and the search is conducted for each query individually (instead of queries in a batch).
All the metrics are measured on every single query and averaged over the whole query set. 

\smallskip\noindent\textbf{Parameter Setting.}
As is suggested by Faiss~\cite{faiss_github}, the number of clusters for IVF is set to be 4,096 as the datasets are at the million-scale. 
For our method, there are two parameters, i.e., $\epsilon_0$ and $B_q$. The theoretical analysis in Section~\ref{subsubsec: construct unbiased estimator} and Section~\ref{subsubsec: quantize query vector} has provided clear suggestions that $\epsilon_0 = \Theta (\sqrt {\log (1/\delta) } )$ and $ B_q = \Theta (\log \log D)$, where $\delta$ is the failure probability. 
In practice, the parameters are fixed to be $\epsilon_0=1.9$ and $B_q=4$ across all the datasets. The empirical parameter study can be found in Section~\ref{subsubsec: verify epsilon0} and Section~\ref{subsubsec: verify BQ}. 
As for the length of the quantization code, it equals to $D$ by definition, but it can also be varied by padding the raw vectors with 0's before generating the quantization codes~\footnote{We emphasize that the padded dimensions will not be retained after the generation, and thus, will not affect the space and time costs related to the raw vectors.}. 
More padded 0's indicate longer quantization codes and higher accuracy due to Theorem~\ref{theorem: estimator} (recall that the error is bounded by $O(1/\sqrt {D})$). By default, the length of the quantization code is set to be the smallest multiple of 64 which is no smaller than $D$ (it is equal to or slightly larger than $D$) in order to make it possible to store the bit string with a sequence of 64-bit unsigned integers.
For the conventional quantization methods (including PQ, OPQ and LSQ), there are two parameters, namely the number of sub-segments of quantization codes $M$ and the number of candidates for re-ranking which should be tuned empirically.
Following the default parameter setting~\cite{scann,faiss_github}, we set the number of partitions to be $M=D/2$. We note that it cannot be further increased as $D$ should be divisible by $M$ for PQ and OPQ.
The number of candidates for re-ranking is varied among 500, 1,000 and 2,500. The experimental results in Section~\ref{subsubsec: time-accuracy trade-off ann search} show that none of the parameters work consistently well across different datasets. For HNSW, we follow its original paper~\cite{malkov2018efficient} by setting 
the number of maximum out-degree of each vertex in the graph as 32 (corresponding to $M_{HNSW}=16$) and a parameter which controls the construction of the graph named $efConstruction$ as 500.

The C++ source codes are compiled by g++ 9.4.0 with \texttt{-Ofast -march=core-avx2} under Ubuntu 20.04LTS. The Python source codes are run on Python 3.8. 
All experiments are run on a machine with AMD Threadripper PRO 3955WX 3.9GHz processor (with Zen2 microarchitecture which supports the SIMD instructions till AVX2) and 64GB RAM. 
The code and datasets are available at 
\url{https://github.com/gaoj0017/RaBitQ}.

\begin{figure*}[th]
  \centering 
    \includegraphics[width=17cm]{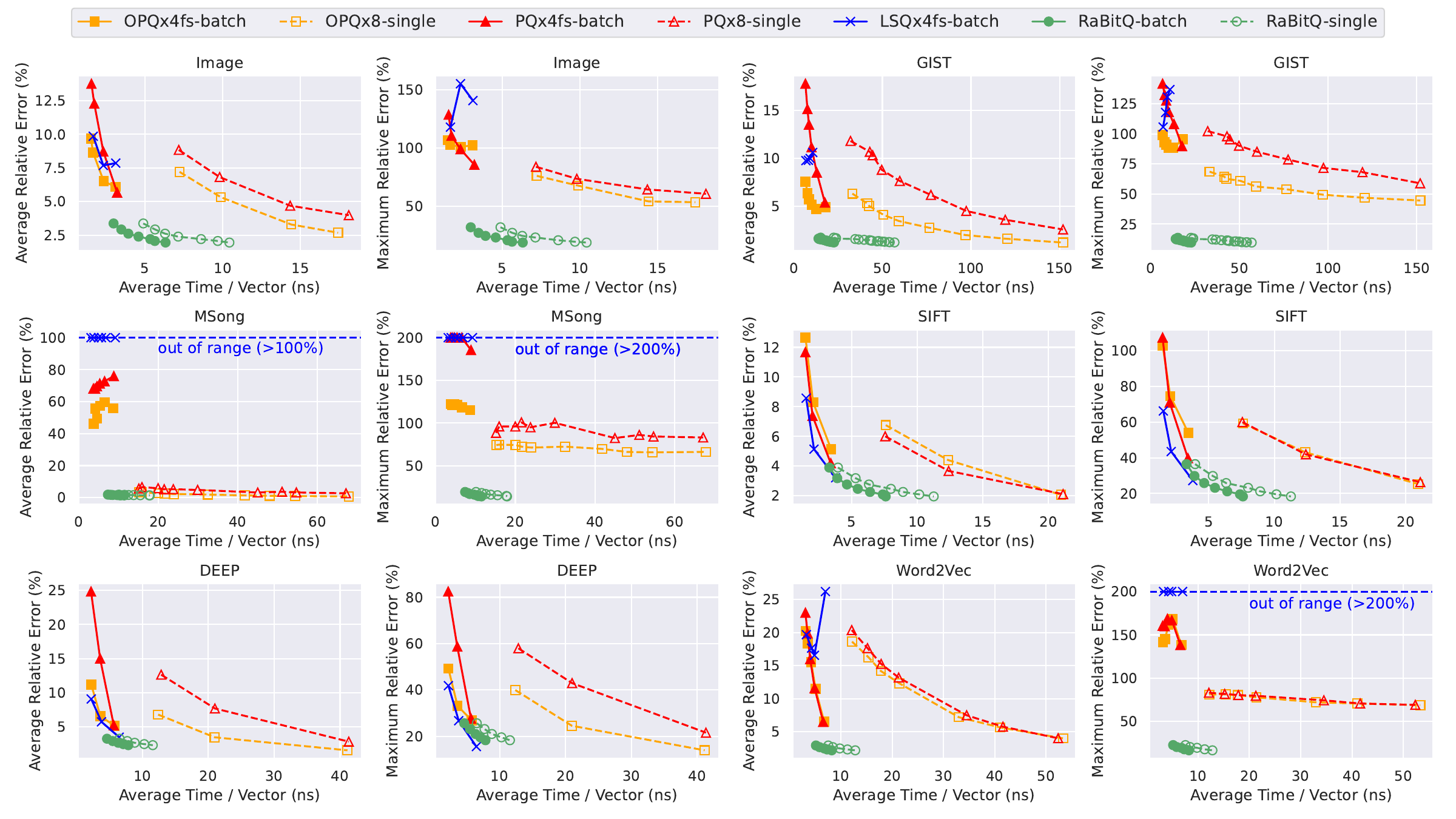}
  \vspace*{-4mm}
  \centering 
  \caption{Time-Accuracy Trade-Off for Distance Approximation. For baseline methods, (1) ``x4fs-batch'' means that the SIMD-based fast implementation is adopted (where 4 bits encode a quantized code and approximate distances for a batch of 32 data vectors are computed each time), and (2) ``x8-single'' means that 8 bits encode a quantized code and the approximate distance of one data vector is computed each time.
  In addition, the results of LSQx8-single are omitted since it, with the implementation from Faiss, has the time cost significantly larger than others.}
  \vspace*{-4mm}
  \label{figure:time-accuracy}
\end{figure*}

\subsection{Experimental Results}
\subsubsection{Time-Accuracy Trade-Off per Vector for Distance Estimation}
\label{subsubsec: time-accuracy trade-off per vector}

We estimate the distance between a data vector (from the set of data vectors) and a query vector (from the set of query vectors) with different quantization methods including PQ, OPQ, LSQ and our RaBitQ method.
We plot the ``average relative error''-``time per vector'' curve (left panels, bottom-left is better) and the ``maximum relative error''-``time per vector'' curve (right panels, bottom-left is better) by varying the length of the quantization codes in Figure~\ref{figure:time-accuracy}.
In particular, for our method, to plot the curve, we vary the length by padding different number of 0's in the vectors when generating the quantization codes. 
For PQ, OPQ and LSQ, we vary the length by setting different $M$ (note that $D$ must be divisible by $M$ for PQ and OPQ). 

Based on the results in Figure~\ref{figure:time-accuracy}, we have the following observations. 
(1) LSQ has much less stable performance than PQ and OPQ. Except for the dataset SIFT and DEEP, LSQx4fs has its accuracy worse than PQx4fs and OPQx4fs. 
(2) Comparing the solid curves, we find that under the default setting of the number of bits (which corresponds to the last point in the red and orange solid curves and the first point in the green solid curve), 
our method shows consistently better accuracy than PQ and OPQ while having comparable efficiency on all the tested datasets. 
We emphasize that in the default setting, the length of the quantization code of our method is only around a half of those of PQ and OPQ (i.e., $D$ v.s. $2D$). 
(3) Comparing the dashed curves, we find that our method has significantly better efficiency than PQ and OPQ when reaching the same accuracy. 
(4) On the dataset Msong, PQx8 and OPQx8 have normal accuracy while PQx4fs and OPQx4fs have disastrous accuracy. 
It indicates that the reasonable accuracy of the conventional quantization methods with $k=8$ does not indicate its normal performance with $k=4$. 
Thus, it is not always feasible to speed up a conventional quantization method with the fast SIMD-based implementation~\cite{fastscanavx2}.
On the other hand, the efficiency of the conventional quantization methods with $k=8$ is hardly comparable with those with $k=4$ on the other datasets. 
It indicates that the recent success of PQ in the in-memory ANN is largely attributed to the fast SIMD-based implementation. Thus, it is not a choice to replace the fast SIMD-based implementation with the original one in pursuit of the stability. 
(5) Except for the dataset SIFT and DEEP, PQx4fs and OPQx4fs have their maximum relative error of around 100\%. 
It indicates that PQ and OPQ do not robustly produce high-accuracy estimated distances even on the datasets they perform well in general.
As a comparison, our method has its maximum relative error at most 40\% on all the tested datasets. 

\subsubsection{Time in the Indexing Phase}
\label{subsubsec: indexing time}
\begin{table}[h]
\vspace{-2mm}
\caption{The Indexing Time for the GIST Dataset}
\label{table:indexing time}
\vspace{-4mm}
\label{tab:indexing}
\begin{tabular}{c|c|c|c|c}
\hline
 & RaBitQ      & PQ &  OPQ & LSQ \\ \hline
Time   & 117s   & 105s    &  291s   & time-out (>24 hours)  \\ \hline
\end{tabular}
\vspace{-2mm}
\end{table}

In Table~\ref{tab:indexing}, we report the indexing time of the quantization methods ($k=4$ for PQ, OPQ and LSQ) under the default parameter setting on the GIST dataset with 32 threads on CPU.
The results show that the indexing time is not a bottleneck for our method, PQ and OPQ since all of them can finish the indexing phase within a few mins. However, for LSQ, it takes more than 24 hours. This is because in LSQ, the step of mapping a data vector to its quantization code is NP-Hard~\cite{lsq,lsq++}.
Although several techniques have been proposed for approximately solving the NP-Hard problem~\cite{lsq,lsq++}, the time cost is still much larger than that of PQ, which largely limits its usage in practice.

\subsubsection{Time-Accuracy Trade-Off for ANN Search}
\label{subsubsec: time-accuracy trade-off ann search}
\begin{figure*}[ht]
  \centering 
    \includegraphics[width=17cm]{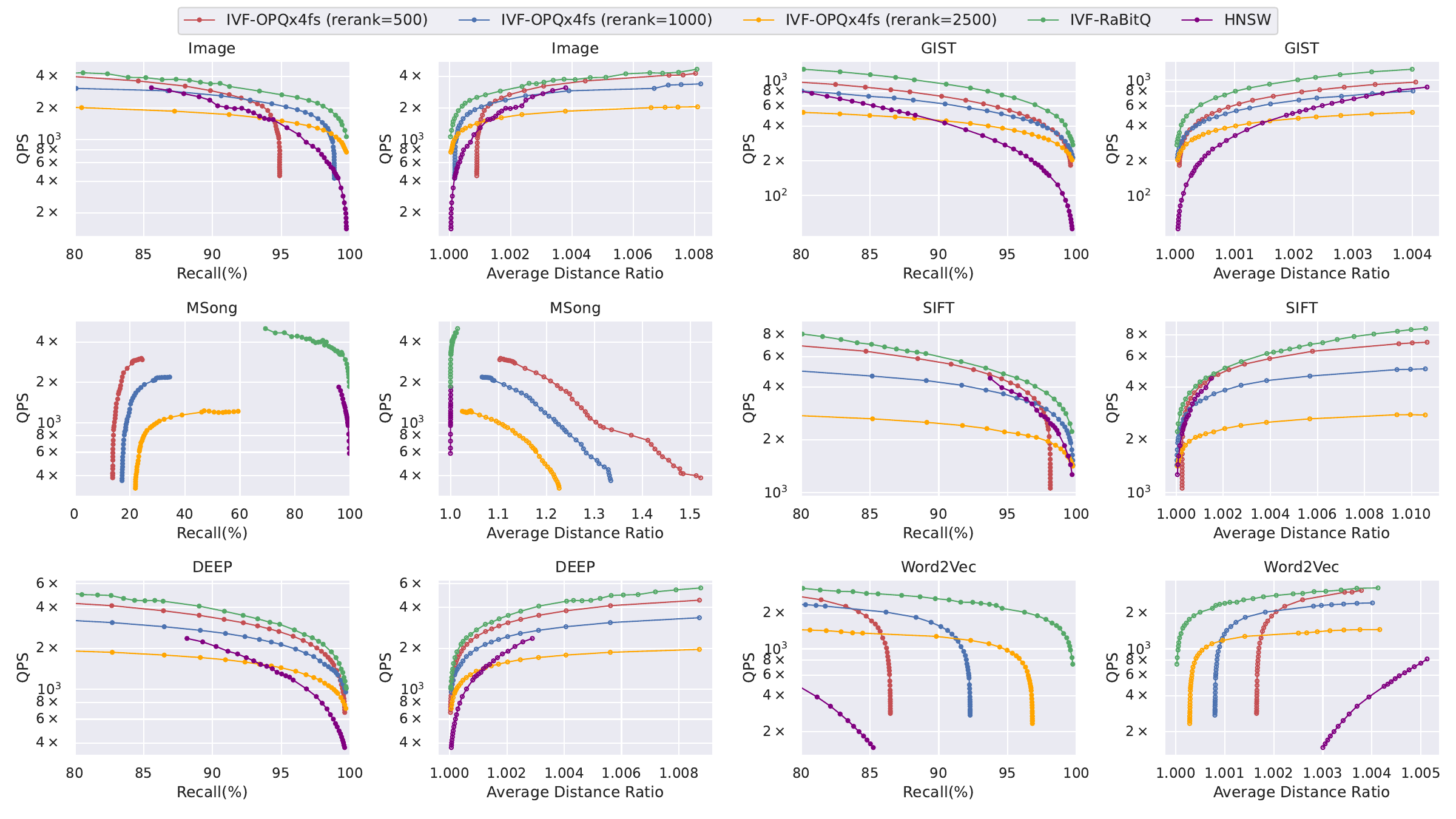}
  \vspace*{-4mm}
  \caption{
  Time-Accuracy Trade-Off for ANN Search. The parameter $rerank$ represents the number of candidates for re-ranking.
  }
  \vspace*{-4mm}
  \label{figure:QPS}
\end{figure*}
We then measure the performance of the algorithms when they are used in combination with the IVF index for ANN search. 
Considering the results in Section~\ref{subsubsec: time-accuracy trade-off per vector}, we only include OPQx4fs-batch and RaBitQ-batch for the comparison as other methods or implementations are in general dominated when the quantization codes are allowed to be packed in batch.
As a reference, we also include HNSW for comparison. We then plot the ``QPS''-``recall'' curve (left panel, upper-right is better) and the ``QPS''-``average distance ratio'' curve (right panel, upper-left is better) by varying the number of buckets to probe in the IVF index for the quantization methods in Figure~\ref{figure:QPS}. The curves for HNSW are plotted by varying a parameter named $efSearch$ which controls the QPS-recall tradeoff of HNSW.
For OPQ, we show three curves which correspond to three different numbers of candidates for re-ranking. 
Based on Figure~\ref{figure:QPS}, we have the following observations. 
(1) On all the tested datasets, our method has consistently better performance than OPQ regardless of the re-ranking parameter. We emphasize that it has been reported that on the datasets SIFT, DEEP and GIST, OPQx4fs has good empirical performance~\cite{fastscanavx2}. Our method also consistently outperforms HNSW on all the tested datasets.
(2) On the dataset MSong, the performance of OPQ is disastrous even with re-ranking applied. In particular, as the IVF index probes more buckets, the recall abnormally decreases because OPQ introduces too much error on the estimated distances. The poor accuracy shown in Figure~\ref{figure:time-accuracy} can explain the disastrous failure. 
(3) No single re-ranking parameter for OPQ works well across all the datasets. On SIFT, DEEP and GIST, 1,000 of candidates for re-ranking suffice to produce a nearly perfect recall while on Image and Word2Vec, a larger number of candidates for re-ranking is needed. We note that the tuning of the re-ranking parameter is often exhaustive as the parameters are intertwined with many factors such as the datasets and the other parameters. Prior to the testing, there is no reliable way to predict the optimal setting of parameters in practice.
In contrast, recall that as is discussed in Section~\ref{subsubsec: construct unbiased estimator} and Section~\ref{subsubsec: quantize query vector}, in our method, the theoretical analysis provides explicit suggestions on the parameters. Thus, our method requires no tuning. 

\begin{figure}[thb]
  \centering 
  \vspace{-2mm}
  \includegraphics[width=\linewidth]{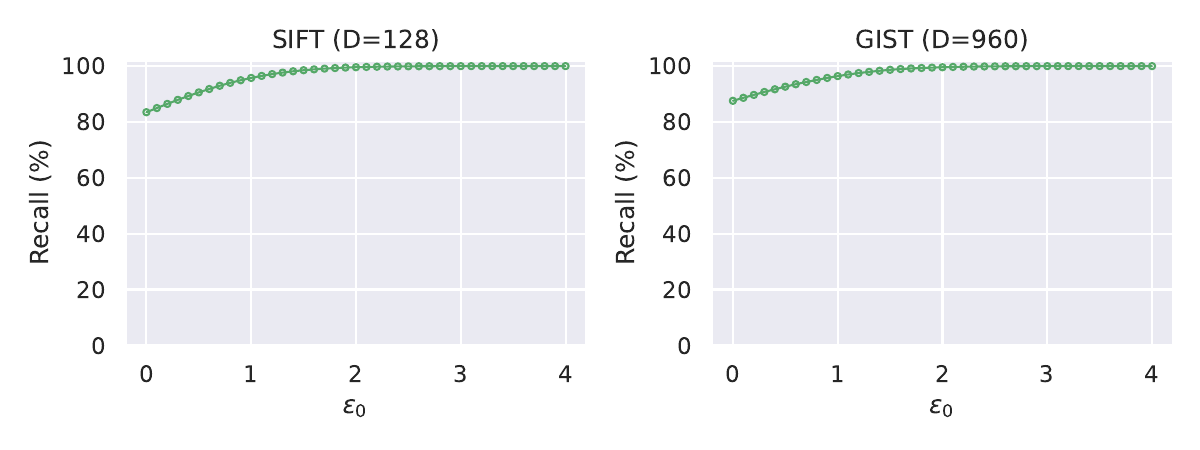}
  \vspace{-4mm}
  \caption{Verification Study on $\epsilon_0$.}
  \vspace{-4mm}
  \label{fig:epsilon}
\end{figure}

\subsubsection{Results for Verifying the Statement about $\epsilon_0$}
\label{subsubsec: verify epsilon0}

$\epsilon_0$ is a parameter which controls the confidence interval of the error bound (see Section~\ref{subsubsec: construct unbiased estimator}). When the RaBitQ method is applied in ANN search, it further controls the probability that we correctly send the NN to re-ranking (see Section~\ref{sec: RaBitQ for ANN}). 
In particular, to make sure the failure probability be no greater than $\delta$, the theoretical analysis in Section~\ref{subsubsec: construct unbiased estimator} suggests to set $\epsilon_0=\Theta (\sqrt {\log (1/\delta)} )$. 
We emphasize that the statement is independent of any other factors such as the datasets or the setting of other parameters. This is the reason that the parameter needs no tuning. 
In Figure~\ref{fig:epsilon}, we provide the empirical verification on the statement. In particular, we plot the ``recall''-``$\epsilon_0$'' curve by varying $\epsilon_0$ from 0.0 to 4.0. 
The recall is measured by estimating the distances for \textit{all} the data vectors and decide the vectors to be re-ranked based on the strategy in Section~\ref{sec: RaBitQ for ANN} (note that if a true nearest neighbor is not re-ranked, it will be missed). 
Thus, the factors (other than the error of quantization methods) which may affect the recall are eliminated.
Figure~\ref{fig:epsilon} shows that on two different datasets, both curves show highly similar trends that it achieves nearly perfect recall at around $\epsilon_0=1.9$.

\begin{figure}[thb]
  \centering 
  \includegraphics[width=\linewidth]{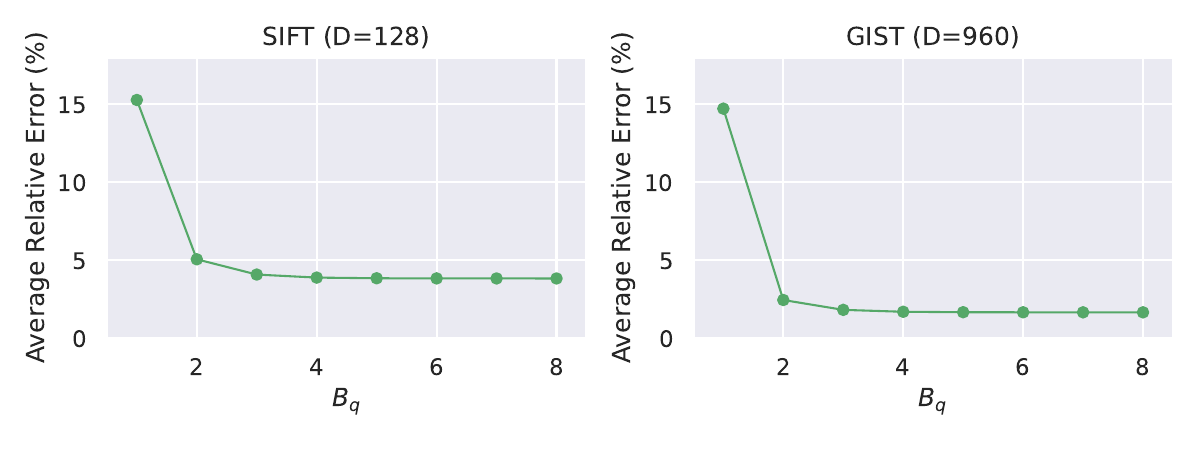}
  \vspace{-4mm}
  \caption{Verification Study on $B_q$.}
  \vspace{-4mm}
  \label{fig:BQ}
\end{figure}
\subsubsection{Results for Verifying the Statement about $B_q$}
\label{subsubsec: verify BQ}

$B_q$ is a parameter which controls the error introduced in the computation of $\left< \mathbf{\bar o}, \mathbf{q} \right> $. 
Due to our analysis in Section~\ref{subsubsec: quantize query vector}, $B_q=\Theta (\log \log D)$ suffices to make sure that the error introduced in the computation of $\left< \mathbf{\bar o}, \mathbf{q} \right> $ is much smaller than the error of the estimator. We note that $\Theta (\log \log D)$ varies extremely slowly with respect to $D$, and thus, it can be viewed as a constant when the dimensionality does not vary largely. 
In Figure~\ref{fig:BQ}, we provide the empirical verification on the statement. In particular, we plot the ``average relative error''-``$B_q$'' curve by varying $B_q$ from 1 to 8.
Figure~\ref{fig:BQ} shows that on two different datasets, both curves show highly similar trends that the error converges quickly at around $B_q=4$.
On the other hand, we would also like to highlight that further reducing $B_q$ would produce unignorable error in the computation of $\left< \mathbf{\bar o}, \mathbf{q} \right> $. 
In particular, when $B_q=1$, i.e., both query and data vectors are quantized into binary strings, the error is much larger than the error when $B_q=4$. 
This result may help to explain why the binary hashing methods cannot achieve good empirical performance. 

\begin{figure}[th]
  \centering 
  \vspace{-2mm}
    \includegraphics[width=\linewidth]{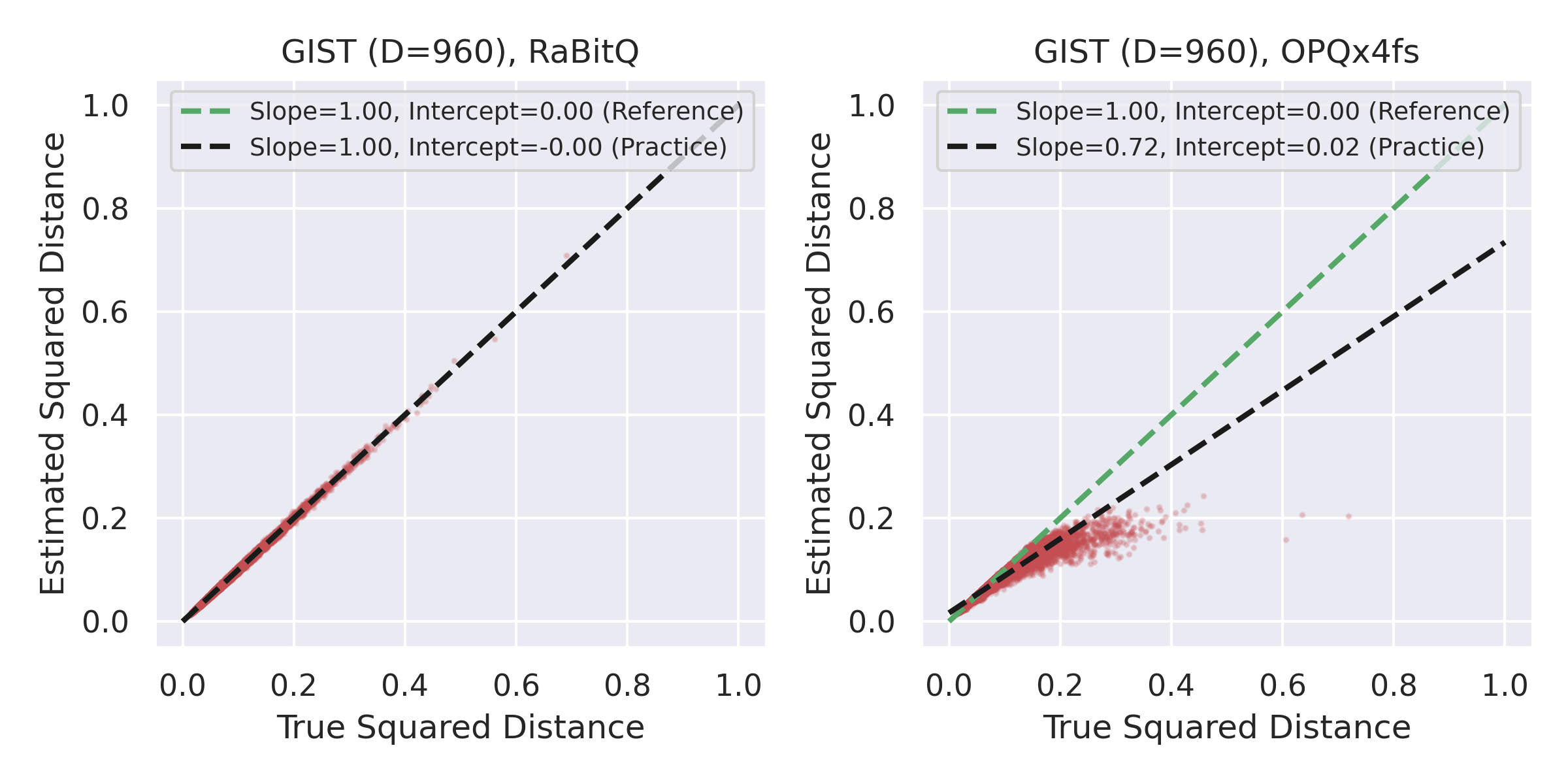}
  \vspace{-4mm}
  \caption{Verification Study for Unbiasedness.}
  \vspace{-4mm}
  \label{figure:unbiasedness}
\end{figure}
\subsubsection{Results for Verifying the Unbiasedness}
\label{subsubsec: verify unbiasedness}
In Figure~\ref{figure:unbiasedness}, we verify the unbiasedness of our method and show the biasedness of OPQ. 
We collect $10^7$ pairs of the estimated squared distances and the true squared distances between the query and data vectors (i.e., the first 10 query vectors in the query set and the $10^6$ data vectors in the full dataset of GIST) to verify the unbiasedness. 
The values of the distances are normalized by the maximum true squared distances. 
We fit the $10^7$ pairs with linear regression and plot the result with the black dashed line. 
We note that if a method is unbiased, the result of the linear regression should have the slope of 1 and the y-axis intercept of 0 (the green dashed line as a reference).
Figure~\ref{figure:unbiasedness} clearly shows that our method is unbiased, which verifies the theoretical analysis in Section~\ref{subsubsec: construct unbiased estimator}. 
On the other hand, the estimated distances produced by OPQ is clearly biased. 
\section{related work}
\label{sec:related work}

\smallskip\noindent\textbf{Approximate Nearest Neighbor Search.} 
Existing studies on ANN search are usually categorized into four types: (1) graph-based methods~\cite{malkov2018efficient, NSW, li2019approximate, fu2019fast, fu2021high, tau_mrng}, (2) quantization-based methods~\cite{jegou2010product, ge2013optimized, scann, ITQ, additivePQ,lsq,lsq++,composite_quantization, vaq}, (3) tree-based methods~\cite{muja2014scalable, dasgupta2008random, beygelzimer2006cover, M_tree} and (4) hashing-based methods~\cite{indyk1998approximate, datar2004locality, c2lsh, tao2010efficient, huang2015query, sun2014srs, lu2020vhp, zheng2020pm, james_cheng, dblsh, idec, lccslsh}. 
We refer readers to recent tutorials~\cite{tutorialThemis, tutorialXiao} and  benchmarks/surveys~\cite{li2019approximate, annbenchmark, graphbenchmark, dobson2023scaling_billion_benchmark,wang2023graph,aumullerrecent2023} for a comprehensive review. 
We note that recently, many studies design algorithms or systems by jointly considering different types of methods so that a method can enjoy the merits of both sides~\cite{lu2021hvs, lsh_apg, sptag, NGT-QG, blink_scalar_quantization_graph,finger,linkandcode}.
Our work proposes a quantization method which provides a sharp error bound and good empirical performance at the same time. 
Just like the conventional quantization methods, it can work as a component in an integrated algorithm or system.
Our method has two additional advantages: (1) it involves no parameter tuning and (2) it supports efficient distance estimation for a single quantization code.
These advantages may further smoothen its combination with other types of methods.
Recently, there are a thread of methods which apply machine learning (ML) on ANN~\cite{learning2route2019ml, reinforcement2route, adaptive2020ml, Dong2020Learning, learning_ann_survey,lider,distill-vq}. 

\smallskip\noindent\textbf{Quantization.} 
There is a vast literature about the quantization of high-dimensional vectors from different communities including machine learning, computer vision and data management~\cite{jegou2010product,additivePQ,ITQ,vaq,composite_quantization,ge2013optimized,lsq,lsq++, vaplusfile, splitvq, vafile}. 
We refer readers to comprehensive surveys~\cite{learningtohash,learningtohashsurvey, ite_matsui_2018_pq_survey, dataseries_benchmark} and reference books~\cite{revision1_bound, revision2_bound}. 
It is worth of mentioning that in~\cite{splitvq}, a quantization method called Split-VQ was mentioned. The method covers the major idea of PQ, i.e., it splits the vectors into sub-segments, constructs sub-codebooks for each sub-segment and forms the codebook with Cartesian product.
Besides PQ and its variants, there are other types of quantization methods, e.g., scalar quantization~\cite{blink_scalar_quantization_graph,vaplusfile,vafile}.
These methods quantize the scalar values of each dimension separately, which often adopt more moderate compression rates than PQ in exchange for better accuracy. 
In particular, VA+ File~\cite{vaplusfile}, a scalar quantization method, has shown leading performance on the similarity search of data series according to a recent benchmark~\cite{dataseries_benchmark}. 
Besides the studies on the quantization algorithms, we note that hardware-aware optimization (with SIMD, GPU, FPGA, etc) also makes significant contributions to the performance of these methods~\cite{fastscan,fastscanavx2,2023wenqi,liu2023juno,johnson2019billion_faiss}. To inherit the merits of the well-developed hardware-aware optimization, in this work, we reduce our computation to the computation of PQ (Section~\ref{subsec: efficient computation}). However, RaBitQ, in its nature, can be implemented with much simpler bitwise operations (which is not possible for PQ and its variants). It remains to be an interesting question whether dedicated hardware-aware optimization can further improve the performance of RaBitQ.

\smallskip\noindent\textbf{Theoretical Studies on High-Dimensional Vectors.} 
The theoretical studies on high-dimensional vectors are primarily about the seminal Johnson-Lindenstrauss (JL) Lemma~\cite{johnson1984extensions}.
It presents that reducing the dimensionality of a vector to $O(\epsilon^{-2}\log (1/\delta))$ suffices to guarantee the error bound of $\epsilon$.
Recent advances improve the JL Lemma in different aspects. 
For example, \cite{larsen2017optimality} proves the optimality of the JL Lemma.
\cite{fftjlt,blockjlt} propose fast algorithms to do the dimension reduction.
We refer readers to a comprehensive survey~\cite{jlintro}.
Our method fits into a recent line of studies~\cite{2017_focs_additive_error, SIAM_JOC_2022_indyk, 2019_ICDT, 2017_nips_quadsketch}, which target to improve the JL Lemma by compressing high-dimensional vectors into short codes.
As a comparison, to guarantee an error bound of $\epsilon$, the JL Lemma requires a vector of $O(\epsilon^{-2}\log (1/\delta))$ \textit{dimensions} while these studies prove that a short code with $O(\epsilon^{-2}\log (1/\delta))$ \textit{bits} would be sufficient. 
In practical terms, we note that although the existing studies achieve the improvement in theory in terms of the space complexity (i.e., the minimum number of bits needed for guaranteeing a certain error bound), they care less about the improvement in efficiency. 
In particular, these methods do not suit the in-memory ANN search because, for estimating the distance during the query phase, they need decompress the short codes and compute the distances with the decompressed vectors, which degrades to the brute force in efficiency. 
For this reason, these methods have not been adopted in practice.
In contrast, our method supports practically efficient computation as is specified in Section~\ref{subsec: efficient computation}.

\smallskip\noindent\textbf{Signed Random Projection.} 
We note that there are a line of studies named signed random projection (SRP) which generate a short code for estimating the angular values between vectors via binarizing the vectors after randomization~\cite{mose_lsh, srp-pmlr-v80-kang18b, srp-pmlr-v180-dubey22a, srp-superbit}.
We note that our method is different from these studies in the following aspects. 
(1) Problem-wise, SRP targets to unbiasedly estimate the angular value while RaBitQ targets to unbiasedly estimate the inner product (and further, the squared distances).
Note that the relationship between the angular value and the inner product is non-linear. 
The unbiased estimator for one does not trivially derive an unbiased estimator for the other. 
(2) Theory-wise, RaBitQ has a stronger type of guarantee than SRP. In particular, RaBitQ guarantees that every data vector has its distance within the bounds with high probability. In contrast, SRP only bounds the variance, i.e., the ``average'' squared error, and it does not provide a bound for every estimated value. Thus, it cannot help with the re-ranking in similarity search. 
(3) Technique-wise, in SRP the bit strings are viewed as some hashing codes while in RaBitQ, the bit strings are the binary representations of bi-valued vectors. 
Moreover, SRP maps both the data and query vectors to bit strings, which introduces error from both sides. 
In contrast, RaBitQ quantizes the data vectors to be bit strings and the query vectors to be vectors of 4-bit unsigned integers. 
Theorem~\ref{theorem: 3.3} proves that quantizing the query vectors 
only introduces negligible error. 
Thus, RaBitQ only introduces the error from the side of the data vector. 

\section{conclusion}
\label{sec:conclusion and discussion}

In conclusion, we propose a novel randomized quantization method RaBitQ which has clear advantages in both empirical accuracy and rigorous theoretical error bound over PQ and its variants. 
The proposed efficient implementations based on simple bitwise operations or fast SIMD-based operations further make it stand out in terms of the time-accuracy trade-off for the in-memory ANN search. 
Extensive experiments on real-world datasets verify both (1) the empirical superiority of our method in terms of the time-accuracy trade-off and (2) the alignment of the empirical performance with the theoretical analysis.
Some interesting research directions include applying our method in other scenarios of ANN search (e.g., with graph-based indexes or on other storage devices~\cite{diskann,spann,pqbf,cxlann}).
Besides, RaBitQ can also be trivially applied to unbiasedly estimate cosine similarity and inner product~\footnote{The cosine similarity of two vectors exactly equals to the inner product of their unit vectors. The inner product of $\mathbf{o}$ and $\mathbf{q}$ can be expressed as $\left< \mathbf{o},\mathbf{q}  \right> =\left< \mathbf{o-c+c},\mathbf{q-c+c}  \right>=\| \mathbf{o-c}\|\cdot \|\mathbf{q-c}\|  \cdot \left< (\mathbf{o-c})/\| \mathbf{o-c}\|, (\mathbf{q-c})/\| \mathbf{q-c}\| \right> + \left< \mathbf{o,c} \right>+ \left< \mathbf{q,c} \right> -\| \mathbf{c}\|^2$, where $\mathbf{c}$ is the centroid of the data vectors, and it reduces to the estimation of inner product between unit vectors as we do in Section~\ref{subsubsec: pre-procession}.}, which further implies its potential in maximum inner product search and neural network quantization.

\section*{Acknowledgements}
We would like to thank the anonymous reviewers for providing constructive feedback and valuable suggestions. This research is supported by the Ministry of Education, Singapore, under its Academic Research Fund (Tier 2 Award MOE-T2EP20221-0013, Tier 2 Award MOE-T2EP20220-0011, and Tier 1 Award (RG77/21)). Any opinions, findings and conclusions or recommendations expressed in this material are those of the author(s) and do not reflect the views of the Ministry of Education, Singapore.

\bibliographystyle{ACM-Reference-Format}
\bibliography{main}

\appendix
\section*{appendix}

\section{The Proof of Lemma~\ref{lemma: geometry}}
\label{proof: geometry}
\begin{proof}
When $\mathbf{o}$ and $\mathbf{q}$ are collinear, i.e., $\mathbf{q}=-\mathbf{o} $ or $\mathbf{q}=\mathbf{o} $, (\ref{eq: collinear}) can be easily verified by definition.
When $\mathbf{o}$ and $\mathbf{q}$ are non-collinear, they can be hosted in a two-dimensional subspace.
We first find a pair of (mutually orthogonal unit) coordinate vectors of the subspace, i.e., $\mathbf{o}$ and $\mathbf{e}_1 := \frac{\mathbf{q}  - \left< \mathbf{q},\mathbf{o}  \right> \mathbf{o}}{\| \mathbf{q}  - \left< \mathbf{q},\mathbf{o}  \right> \mathbf{o}\|}$, which can be verified by 
\begin{align}
    \left< \mathbf{o} , \mathbf{e}_1  \right> =&\left< \mathbf{o}, \frac{\mathbf{q}  - \left< \mathbf{q},\mathbf{o}  \right> \mathbf{o}}{\| \mathbf{q}  - \left< \mathbf{q},\mathbf{o}  \right> \mathbf{o}\|}\right> = \frac{\left< \mathbf{q,o} \right> -\left< \mathbf{q,o} \right> \cdot \| \mathbf{o}\|^2 }{\| \mathbf{q} - \left< \mathbf{q}, \mathbf{o}  \right> \mathbf{o}\|  } =0 
\end{align}    
We next decompose $\mathbf{\bar o}$ and $\mathbf{q} $ based on the coordinate vectors $\mathbf{o}$ and $\mathbf{e}_1$ as follows.
\begin{align}
    \mathbf{\bar o}= &\left( \mathbf{\bar o} - \left< \mathbf{\bar o}, \mathbf{o}  \right> \mathbf{o}  -\left< \mathbf{\bar o}, \mathbf{e}_1  \right> \mathbf{e}_1 \right)  + \left< \mathbf{\bar o}, \mathbf{o}  \right> \mathbf{o} + \left< \mathbf{\bar o}, \mathbf{e}_1  \right> \mathbf{e}_1   
    \\ \mathbf{q} = &\left< \mathbf{q}, \mathbf{o}  \right> \mathbf{o} + \left< \mathbf{q}, \mathbf{e}_1  \right> \mathbf{e}_1   \label{eq: decompose q}
\end{align}
where (\ref{eq: decompose q}) is because $\mathbf{q}$ is in the subspace. Then because $\left( \mathbf{\bar o} - \left< \mathbf{\bar o}, \mathbf{o}  \right> \mathbf{o}  -\left< \mathbf{\bar o}, \mathbf{e}_1  \right> \mathbf{e}_1 \right)$ is orthogonal to the subspace and $\mathbf{o} \perp \mathbf{e}_1$, we have
\begin{align}
    \left< \mathbf{\bar o}, \mathbf{q}  \right> = &\left< \mathbf{\bar o}, \mathbf{o} \right> \cdot \left< \mathbf{o}, \mathbf{q}  \right> +   \left< \mathbf{\bar o}, \mathbf{e}_1  \right> \cdot  \left< \mathbf{q}, \mathbf{e}_1  \right>  \label{eq: orthogonal}
    \\ = & \left< \mathbf{\bar o}, \mathbf{o} \right> \cdot \left< \mathbf{o}, \mathbf{q} \right> +   \left< \mathbf{\bar o}, \mathbf{e}_1  \right> \cdot  \sqrt {1- \left< \mathbf{o}, \mathbf{q} \right>^2} \label{eq: simple computation}
\end{align}
where (\ref{eq: simple computation}) is due to the Pythagorean Theorem.
\end{proof}

\section{The Analysis for the Concentration Phenomenon}
\label{appendix: concentration}
In this part, we provide rigorous analysis for the concentration phenomenon presented in Section~\ref{subsubsec: geometric}. In particular, we will analyze the expected value of $\left< \mathbf{\bar o}, \mathbf{o}  \right> $ (Section~\ref{appendix: expected value of o bar}), the extent of concentration of $\left< \mathbf{\bar o}, \mathbf{o}  \right> $ (Section~\ref{appendix: concentration of o bar}) and the joint distribution of $(\left< \mathbf{\bar o},\mathbf{o}  \right>, \left< \mathbf{\bar o},\mathbf{e}_1  \right>)$ (Section~\ref{appendix: joint distribution}). We summarize the conclusions in Lemma~\ref{lemma: joint distribution} and empirically verify them in Figure~\ref{fig:empirical verification of lemma 3.2}.

\subsection{The Expected Value of $\left< \mathbf{\bar o}, \mathbf{o}  \right> $}
\label{appendix: expected value of o bar}
As is analyzed in Section~\ref{subsubsec: geometric}, $\left< \mathbf{\bar o}, \mathbf{o}  \right> $ and $\left< \mathbf{\bar o}, \mathbf{e}_1  \right> $ correspond to the projection of the random vector $\mathbf{\bar o}$ onto two mutually orthogonal directions. 
In order to analyze the joint distribution of the random variables, let us first revisit the process of the generation of $\mathbf{\bar o}$. 
The generation of $\mathbf{\bar o}$ involves two steps. 
\underline{First}, the algorithm randomly transforms a deterministic codebook $\mathcal{C} $ into $\mathcal{C}_{rand}$ with a random orthogonal transformation $P$.
\underline{Second}, it chooses the vector $\mathbf{\bar o} $ which has the largest inner product with $\mathbf{o} $ from the vectors in $\mathcal{C}_{rand}$. 
We next deduce from the definition (generation) of $\left< \mathbf{\bar o}, \mathbf{o}  \right> $ as follows.
\begin{align}
    \left< \mathbf{\bar o}, \mathbf{o}  \right>   =&  \max_{\mathbf{x} \in C } \left< P\mathbf{x}, \mathbf{o}  \right>   \label{eq: obar by definition}
    \\ =& \max_{\mathbf{x} \in C } \left< \mathbf{x}, P^{-1}\mathbf{o}  \right> = \max_{\mathbf{x} \in C } \sum_{i=1}^{D} \mathbf{x}[i]\cdot ( P^{-1}\mathbf{o})[i]
    \label{eq: obar invariant to rotation}
    \\ =&\frac{1}{\sqrt {D} } \sum_{i=1}^{D}  \left| (P^{-1}\mathbf{o})[i] \right| = \frac{1}{\sqrt {D} } \| P^{-1}\mathbf{o}\|_{\ell_1} \label{eq: definition of l1 norm}
\end{align}
where (\ref{eq: obar by definition}) is due to the process of generation of $\mathbf{\bar o} $. (\ref{eq: obar invariant to rotation}) is because the inner product is invariant to orthogonal transformation (rotation). (\ref{eq: definition of l1 norm}) is due to the definition of our codebook $\mathcal{C}$ and the definition of $\ell_1$ norm. Specifically, as is analyzed in Section~\ref{subsubsec: quantization code}, the entry of $\mathbf{x} \in \mathcal{C} $ can only be $1/\sqrt {D}$ or $-1/\sqrt {D}$. To maximize the inner product, we only need to pick the $\mathbf{x}$ which has its signs of the entries match the vector $P^{-1}\mathbf{o}$. Thus, the result of the inner product is the summation the absolute values as is presented in (\ref{eq: definition of l1 norm}), where $\| \cdot \|_{\ell_1}$ is the $\ell_1$ norm.

We note that $\mathbf{o}$ is a unit vector. $P$ is a random orthogonal transformation matrix (i.e., random rotation), whose inverse matrix (inverse rotation) is also a random orthogonal transformation matrix. Thus, the vector $P^{-1}\mathbf{o}$ follows the uniform distribution on the unit sphere $\mathbb{S}^{D-1} $ in the $D$-dimensional space $\mathbb{R}^D$. We note that the distribution is well studied~\cite{uniform_spherical_distribution,vershynin_2018}. We restate some conclusions about the distribution with the following lemma.
\begin{lemma}(\cite{uniform_spherical_distribution, vershynin_2018})
\label{lemma: distribution}
    For a $D$-dimensional random vector $\mathbf{x}=(\mathbf{x}[1], \mathbf{x}[2],...,\mathbf{x}[D]) $ which follows the uniform distribution on the unit sphere, the probability density function of its every coordinate $\mathbf{x}[1], \mathbf{x}[2], ..., \mathbf{x}[D]$ is given as 
    \begin{align}
        p_D(x)=\frac{\Gamma (\frac{D}{2} )}{\sqrt {\pi} \Gamma (\frac{D-1}{2} ) } (1-x^2)^{\frac{D-3}{2} }, x\in [-1,1]
    \end{align}
    where $\Gamma(\cdot)$ is the Gamma function. The tail bound is given as 
    \begin{align}
        \mathbb{P}\left\{ \left| \mathbf{x}[i]\right| > \frac{t}{\sqrt {D} }     \right\}  \le 2\exp \left( -c_0 t^2 \right) 
    \end{align}
    where $c_0$ is a constant, $i=1,2,...,D$.
\end{lemma}
Based on the explicit expression of the probabilistic density function, we next derive the expected value of $\left< \mathbf{\bar o},\mathbf{o}  \right> $ as follows. 
\begin{align}
    \mathbb{E} \left[ \left< \mathbf{\bar o}, \mathbf{o}  \right>  \right]  
    =& \frac{1}{\sqrt {D} } \cdot \mathbb{E} \left[ \sum_{i=1}^{D}  \left| (P^{-1}\mathbf{o})[i] \right| \right] \label{eq: 31}
    \\=& \sqrt {D} \cdot \mathbb{E} \left[ \left| (P^{-1}\mathbf{o})[1] \right| \right]  \label{eq: 32}
    \\=& \sqrt {\frac{D}{\pi}} \frac{\Gamma (\frac{D}{2} )}{\Gamma (\frac{D-1}{2} ) }    \cdot \int_{-1}^1 (1-x^2)^{\frac{D-3}{2} } |x| \mathrm{d}x \label{eq: 33}
    \\=& \sqrt {\frac{D}{\pi}} \frac{2\Gamma (\frac{D}{2} )}{(D-1)\Gamma (\frac{D-1}{2} ) } \label{eq: 37}
\end{align}
where (\ref{eq: 31}) is due to (\ref{eq: definition of l1 norm}).  (\ref{eq: 32}) is due to the linearity of expectation. (\ref{eq: 33}) is due to Lemma~\ref{lemma: distribution}. (\ref{eq: 37}) is by elementary calculus. 

We note that although the expected value of $ \left< \mathbf{\bar o}, \mathbf{o}  \right>$ has a complicated form, i.e., (\ref{eq: 37}), its numerical value is highly stable. When $D$ ranges from $10^2$ to $10^6$, the value ranges from 0.798 to 0.800, which is verified by the observations in Section~\ref{subsubsec: geometric} perfectly.

\subsection{The Concentration of $\left< \mathbf{\bar o}, \mathbf{o}  \right> $}
\label{appendix: concentration of o bar}
We next analyze the extent of the concentration of $\left< \mathbf{\bar o}, \mathbf{o} \right> $. Recall that as is shown in (\ref{eq: definition of l1 norm}), $\left< \mathbf{\bar o},\mathbf{o}  \right> = \frac{1}{\sqrt {D} } \| P^{-1} \mathbf{o}\|_{\ell_1}  $. Let $f(\mathbf{x} ) := \frac{1}{\sqrt {D}} \| \mathbf{x}  \|_{\ell_1}$.
Then $\left< \mathbf{\bar o},\mathbf{o}  \right> = f(P^{-1} \mathbf{o})$. 
We note that $f(\mathbf{x} )$ is a Lipschitz function with the Lipschitz constant of 1, i.e.,
\begin{align}
    \left| f(\mathbf{x})-f(\mathbf{y}) \right|  \le 1\cdot \| \mathbf{x-y}\|
\end{align}
for every $\mathbf{x,y} $ on the unit sphere. 
\begin{proof}
    \begin{align}
        &\left| f(\mathbf{x} )-f(\mathbf{y})  \right| 
        = \frac{1}{\sqrt{D}} \left| \|\mathbf{x}\|_{\ell_1} - \|\mathbf{y}\|_{\ell_1}   \right|  \label{eq: 40 by definition}
        \\ \le &  \frac{1}{\sqrt{D}} \| \mathbf{x-y} \|_{\ell_1} 
        =  \frac{1}{\sqrt{D}} \sum_{i=1}^{D} 1 \cdot |\mathbf{x}[i]- \mathbf{y}[i]| \label{eq: 41 by triangle's inequality}
        \\ \le & \frac{1}{\sqrt{D}}\cdot \sqrt {\sum_{i=1}^{D} 1^2} \cdot \sqrt {\sum_{i=1}^{D} (\mathbf{x}[i]- \mathbf{y}[i])^2 } = \| \mathbf{x-y}\| \label{eq: 42 by cauchy}
    \end{align}
    where (\ref{eq: 40 by definition}) is by definition. (\ref{eq: 41 by triangle's inequality}) is by triangle's inequality. (\ref{eq: 42 by cauchy}) is due to Cauchy-Schwarz inequality.
\end{proof}
Recall that $P^{-1} \mathbf{o} $ is a random vector which follows the uniform distribution on the unit sphere. 
There is a well-known lemma~\cite{vershynin_2018} which presents that passing a random vector which follows the uniform distribution on the unit sphere through a Lipschitz function produces a highly concentrated distribution. 
The specific result is presented as follows.
\begin{lemma} (\cite{vershynin_2018})
    Let $\mathbf{x} $ be a $D$-dimensional random vector which follows the uniform distribution on the unit sphere, $f(\mathbf{x}) $ is a Lipschitz function with the Lipschitz constant of $L$. Then 
    \begin{align}
        \mathbb{P} \left\{ \left| f(\mathbf{x}) -\mathbb{E} \left[ f(\mathbf{x} ) \right] \right|  \ge t    \right\}  \le 2\exp \left( - \frac{cDt^2}{L^2}  \right) 
    \end{align}
    where $c$ is a constant.
\end{lemma}
Plugging in our $f(\mathbf{x})$ immediately yields the following result.
\begin{align}
    \mathbb{P} \left\{ \left| \left< \mathbf{\bar o},\mathbf{o}  \right>  -\mathbb{E} \left[ \left< \mathbf{\bar o},\mathbf{o}  \right> \right] \right|  \ge t \right\}  \le 2\exp \left( - cDt^2 \right) 
    \\\mathbb{P} \left\{ \left| \left< \mathbf{\bar o},\mathbf{o}  \right>  -\mathbb{E} \left[ \left< \mathbf{\bar o},\mathbf{o}  \right> \right] \right|  \ge \frac{u}{\sqrt {D} }     \right\}  \le 2\exp \left( - cu^2 \right) \label{eq: let epsilon0=t over sqrt D}
\end{align}
where (\ref{eq: let epsilon0=t over sqrt D}) is by letting $u=t\sqrt {D}$. The conclusion shows that $\left< \mathbf{\bar o}, \mathbf{o}  \right> $ is highly concentrated around its expectation. It will not deviate from its expectation by $\Omega(1/\sqrt {D} )$ with high probability.

\subsection{The Joint Distribution of $(\left< \mathbf{\bar o},\mathbf{o}  \right>, \left< \mathbf{\bar o},\mathbf{e}_1  \right> )$}
\label{appendix: joint distribution}
We next analyze the distribution of $\left< \mathbf{\bar o}, \mathbf{e}_1  \right>$. Recall that the randomness of both $\left< \mathbf{\bar o}, \mathbf{o}  \right>$ and $\left< \mathbf{\bar o}, \mathbf{e}_1  \right>$ is due to the randomness of $\mathbf{\bar o} $ (which is further due to the randomness of $P$). 
These random variables are correlated with each other, which is undesirable for subsequent analysis.
We first consider decorrelating the random variables by representing $\left< \mathbf{\bar o}, \mathbf{e}_1\right> $ with a function of $\left<\mathbf{\bar o}, \mathbf{o} \right> $ and a random variable which is independent to $\left<\mathbf{\bar o}, \mathbf{o} \right> $. 
In particular, we will show that the joint distribution of $(\left< \mathbf{\bar o},\mathbf{o}  \right> , \left< \mathbf{\bar o},\mathbf{e}_1  \right>)$ is identical to that of 
\begin{align}
    \left( \left< \mathbf{\bar o}, \mathbf{o}  \right>,  \sqrt {1 - \left< \mathbf{\bar o}, \mathbf{o}  \right>^2} \cdot X_1 \right) 
\end{align}
where $X_1$ follows the distribution of $p_{D-1}$ in Lemma~\ref{lemma: distribution}
and is independent to $\left< \mathbf{\bar o}, \mathbf{o}  \right>$. 

\begin{proof}
    Let $\mathbf{u}_1=(1,0,0,...,0), \mathbf{u}_2 = (0,1,0,...,0)$.
    For $\mathbf{o}$ and $\mathbf{e}_1$ where $\mathbf{o} \perp \mathbf{e}_1$, there exists an orthogonal matrix $U$ to align them on $\mathbf{u}_1$ and $\mathbf{u}_2$, i.e., $U \mathbf{o}=\mathbf{u}_1, U \mathbf{e}_1=\mathbf{u}_2$. 
    Then by applying the orthogonal matrix $U$ to both sides of the inner products, we have $\mathbf{\bar x}=\argmax_{\mathbf{x} \in  \mathcal{C}} \left< UP \mathbf{x}, \mathbf{u}_1\right> $,
    $\left< P\mathbf{\bar x}, \mathbf{o} \right> = \left< UP\mathbf{\bar x}, \mathbf{u}_1 \right>$ and $\left< P\mathbf{\bar x}, \mathbf{e}_1 \right> = \left< UP\mathbf{\bar x}, \mathbf{u}_2 \right>$. Thus, to prove the original statement, it is equivalent to prove that the joint distribution of $(\left< UP \mathbf{\bar x}, \mathbf{u}_1  \right>, \left< UP \mathbf{\bar x}, \mathbf{u}_2  \right>)$ is identical to that of 
    \begin{align}
        \left( \left< UP \mathbf{\bar x}, \mathbf{u}_1  \right>,  \sqrt {1 - \left< UP \mathbf{\bar x}, \mathbf{u}_1  \right>^2} \cdot X_1 \right) 
    \end{align}
    where $X_1$ is independent to $\left< UP \mathbf{\bar x}, \mathbf{u}_1  \right> $ and $X_1 \sim p_{D-1}$.
    Because the random orthogonal transformation is rotation-invariant, i.e., the distribution of $UP$ is identical to the distribution of $P$, we can substitute all the $UP$ with $P$. 
    Then the statement reduces to the same form as the original one while replacing $\mathbf{o}$ and $\mathbf{e}_1$ with $\mathbf{u}_1 $ and $\mathbf{u}_2 $, respectively. Thus, in order to prove the statement for a general pair of $\mathbf{o}$ and $\mathbf{e}_1$, it suffices to prove the case of $\mathbf{o}=\mathbf{u}_1$ and $\mathbf{e}_1=\mathbf{u}_2$ without loss of generality. We next prove the case of $\mathbf{u}_1,\mathbf{u}_2$.

    We consider analyzing the distribution based on the Principle of Deferred Decision~\cite{randomized_algorithm_motwani_raghavan_1995}. 
    The basic idea of the Principle of Deferred Decision is that 
    for a randomized algorithm which needs sample several random numbers, we assume that the sampling operation happens at the time when the algorithm accesses the sampled numbers instead of happening in the very beginning.
    In our case, recall that we will sample a random orthogonal matrix $P$.
    Its generation involves sampling $D\times D$ of standard Gaussian random variables as its entries and orthonormalizing the matrix with the Gram–Schmidt orthonormalization.
    We note that the Gram–Schmidt orthonormalization proceeds row by row. 
    In particular, assuming that the first $i-1$ rows of $P$, i.e., $\mathbf{p}_1,...,\mathbf{p}_{i-1}$, have been orthonormalized (i.e., $\mathbf{p}_j \perp \mathbf{p}_{k}, \| \mathbf{p}_j\| =\| \mathbf{p}_k\|=1  , \forall 1\le j < k \le i-1$), we orthonormalize the first $i$ row by letting $\mathbf{p}_i$ be 
    \begin{align}
        \mathbf{p}_i = \frac{\mathbf{g}-\sum_{j=1}^{i-1} \left< \mathbf{g}, \mathbf{p}_j  \right> \mathbf{p}_j}{\left\| \mathbf{g}-\sum_{j=1}^{i-1} \left< \mathbf{g}, \mathbf{p}_j  \right> \mathbf{p}_j\right\|}  
    \end{align}
    where the entries of $\mathbf{g}$ are sampled from a standard random Gaussian distribution.
    Thus, due to the process of Gram–Schmidt orthonormalization, the sampling process can be viewed as a sequential process of $D$ steps where in each step we sample a new row.

    In our algorithm, we note that the joint distribution of $(\left< P \mathbf{\bar x}, \mathbf{u}_1  \right>, \left< P \mathbf{\bar x}, \mathbf{u}_2  \right>)$ depends only on the first two row of $P$. 
    Let us first sample the first row of $P$, i.e., $\mathbf{p}_1$. Then $ \mathbf{\bar x}$ is determined because
    \begin{align}
        \mathbf{\bar x}&=\argmax_{\mathbf{x\in \mathcal{C} } } \left< P \mathbf{x}, \mathbf{u}_1  \right>=\argmax_{\mathbf{x\in \mathcal{C} } } \left<  \mathbf{x}, P^{-1}\mathbf{u}_1  \right> \\&=\argmax_{\mathbf{x\in \mathcal{C} } } \left<  \mathbf{x}, P^{\top}\mathbf{u}_1  \right> 
        = \argmax_{\mathbf{x\in \mathcal{C}}} \left< \mathbf{x}, \mathbf{p}_1  \right>
        \label{appendix eq: inverse of orthogonal matrix}
    \end{align}
    where (\ref{appendix eq: inverse of orthogonal matrix}) is because the inverse of an orthogonal matrix equals to its transpose. 
    For the fixed $\mathbf{\bar x} $, we next analyze the distribution of $\left< P\mathbf{\bar x}, \mathbf{u}_2 \right> $. We note that similarlly, it depends only on the first two rows of $P$ because $\left< P \mathbf{\bar x}, \mathbf{u}_2  \right> = \left< \mathbf{\bar x}, \mathbf{p}_2  \right>$ (recall that $\mathbf{\bar x} $ depends on $\mathbf{p}_1 $).
    For the vector $\mathbf{p}_1 $ and $\mathbf{\bar x} $, there exists an orthogonal matrix $V$ to align them on $\mathbf{v}_1= (1,0,0,...,0)$ and $\mathbf{v}_2= (\left< \mathbf{\bar x}, \mathbf{p}_1 \right>,\sqrt {1-\left< \mathbf{\bar x}, \mathbf{p}_1 \right>^2} ,0,...,0)$, i.e., $\mathbf{v}_1=V \mathbf{p}_1 $ and $\mathbf{v}_2=V \mathbf{\bar x} $.
    Then 
    \begin{align}
        \left< \mathbf{\bar x}, \mathbf{p}_2   \right> &= \left< \mathbf{\bar x}, \frac{\mathbf{g}- \left< \mathbf{g},\mathbf{p_1}  \right>\mathbf{p_1} }{\| \mathbf{g}- \left< \mathbf{g},\mathbf{p_1}  \right>\mathbf{p_1}\| }    \right> \label{appendix eq: definition by gram-schmidt}
        \\&= \left< V\mathbf{\bar x}, \frac{V\mathbf{g}- \left< V\mathbf{g},V\mathbf{p_1}  \right>V\mathbf{p_1} }{\| V\mathbf{g}- \left< V\mathbf{g},V\mathbf{p_1}  \right>V\mathbf{p_1}\| }    \right>\label{appendix eq: 49}
        \\&= \left< \mathbf{v}_2, \frac{\mathbf{g}- \left< \mathbf{g},\mathbf{v_1}  \right>\mathbf{v_1} }{\| \mathbf{g}- \left< \mathbf{g},\mathbf{v_1}  \right>\mathbf{v_1}\| }    \right>\label{appendix eq: 50}
    \end{align}
    where (\ref{appendix eq: definition by gram-schmidt}) is by Gram-Schmidt orthonormalization. (\ref{appendix eq: 49}) is because inner product and Euclidean distance are invariant to orthogonal transformation. (\ref{appendix eq: 50}) is because standard Gaussian random vector is rotational-invariant~\cite{vershynin_2018}, i.e., $V \mathbf{g}$ and $\mathbf{g}$ are identically distributed. 
    We note that $\mathbf{v}_1 $ only has its first entry non-zero. 
    Thus, $\mathbf{g}- \left< \mathbf{g},\mathbf{v_1}  \right>\mathbf{v_1}$ has the first dimension of 0 and 
    has its remaining $D-1$ dimensions to be independent standard Gaussian variables. After normalization, the remaining $D-1$ dimensions follow the uniform distribution on unit sphere in the $(D-1)$-dimensional space and are independent to $\mathbf{p}_1$.
    Recall that $\mathbf{v}_2= (\left< \mathbf{p}_1,\mathbf{\bar x} \right>,\sqrt {1-\left< \mathbf{p}_1,\mathbf{\bar x} \right>^2} ,0,...,0)$. 
    Thus, $\left< \mathbf{\bar x}, \mathbf{p}_2  \right> = \sqrt {1-\left< \mathbf{p}_1,\mathbf{\bar x} \right>^2} \cdot X_1$ where $X_1$ follows the distribution of $p_{D-1}$.
\end{proof}

We summarize our conclusions about the distribution of $(\left< \mathbf{\bar o}, \mathbf{o}  \right> , \left< \mathbf{\bar o}, \mathbf{e}_1  \right> )$ with the following lemma. 
\begin{lemma}[Distribution]
\label{lemma: joint distribution}
Let $\mathbf{o}$ and $\mathbf{e}_1$ be two unit vectors, where $\mathbf{o} \perp \mathbf{e}_1$. 
Let $P$ be a random orthogonal transformation matrix, $\mathcal{C}$ be our constructed deterministic codebook and $\mathbf{\bar x}=\argmax_{\mathbf{x} \in  \mathcal{C}} \left< P \mathbf{x}, \mathbf{o}\right>$, $\mathbf{\bar o}=P \mathbf{\bar x}  $. Then the joint distribution of $(\left< \mathbf{\bar o}, \mathbf{o}  \right>, \left< \mathbf{\bar o}, \mathbf{e}_1  \right>)$ is identical to that of 
\begin{align}
    \left( \left< \mathbf{\bar o}, \mathbf{o}  \right>,  \sqrt {1 - \left< \mathbf{\bar o}, \mathbf{o}  \right>^2} \cdot X_1 \right) \label{eq: lemma3.2 correlation}
\end{align}
where $X_1$ is independent to $\left< \mathbf{\bar o}, \mathbf{o}  \right> $ and $X_1 \sim p_{D-1}$.
The expected value of $\left< \mathbf{\bar o}, \mathbf{o}\right>$ is 
\begin{align}
    \mathbb{E}\left[ \left< \mathbf{\bar o}, \mathbf{o}\right>   \right] = \sqrt {\frac{D}{\pi} } \cdot \frac{2\Gamma(\frac{D}{2} )}{(D-1)\Gamma (\frac{D-1}{2} )}  \label{eq: lemma3.2 where to concentrate}
\end{align}
where $\Gamma(\cdot)$ is the Gamma function. Its concentration bound is 
\begin{align}
    \mathbb{P} \left\{ \left| \left< \mathbf{\bar o}, \mathbf{o}\right> -\mathbb{E}\left[ \left< \mathbf{\bar o}, \mathbf{o}\right>  \right]   \right| > \frac{u}{\sqrt {D} }   \right\} \le 2  \exp \left( -cu^2 \right) \label{eq: lemma3.2 how concentrated}
\end{align}
where $c$ in a constant. 
\end{lemma}

\begin{figure}[thb]
    \centering
    \includegraphics[width=\linewidth]{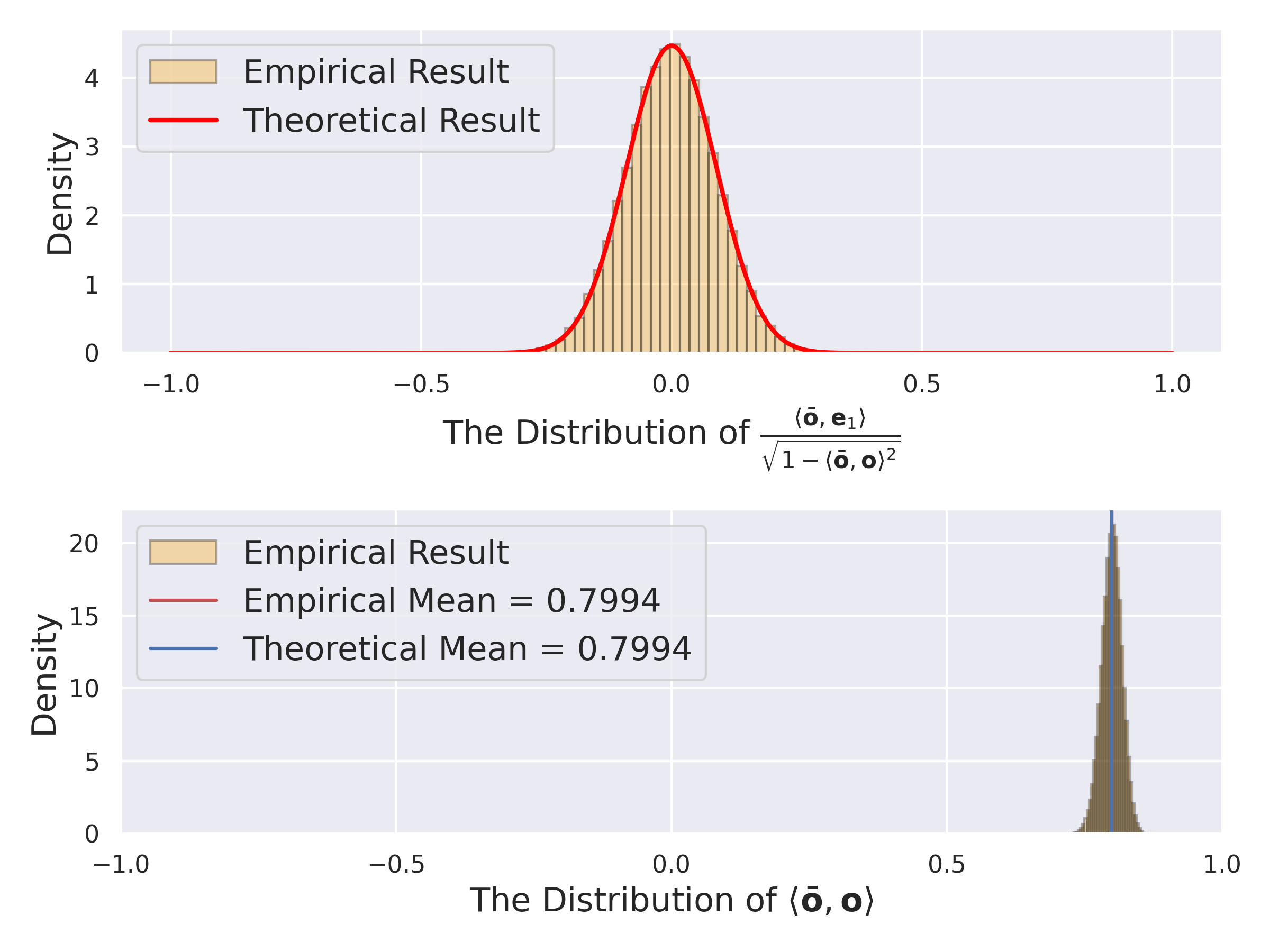}
    \vspace{-8mm}
    \caption{Verification of Lemma~\ref{lemma: joint distribution}.}
    \label{fig:empirical verification of lemma 3.2}
    \vspace{-4mm}
\end{figure}

We next provide empirical verification for the theorem in Figure~\ref{fig:empirical verification of lemma 3.2}.
\underline{First}, in the upper panel of Figure~\ref{fig:empirical verification of lemma 3.2}, the orange histogram represents the empirical distribution of $\frac{\left< \mathbf{\bar o}, \mathbf{e}_1  \right> }{\sqrt {1 - \left< \mathbf{\bar o}, \mathbf{o}  \right>^2 } } $ based on the $10^5$ samples of $P$ in Section~\ref{subsubsec: geometric}. 
Due to (\ref{eq: lemma3.2 correlation}), it follows the distribution of $p_{D-1}$.
The red curve plots the theoretical density function of $p_{D-1}$ as is specified in Lemma~\ref{lemma: distribution}.
It shows that the empirical results and the theoretical results match perfectly, which verifies the correctness of our analysis.
\underline{Second}, in the lower panel of Figure~\ref{fig:empirical verification of lemma 3.2}, the orange histogram represents the empirical distribution of $\left< P \mathbf{\bar x}, \mathbf{o} \right> $ based on the aforementioned $10^5$ samples. It shows that $\left< P\mathbf{\bar x} , \mathbf{o} \right> $ is indeed highly concentrated around its mean value and the empirical mean value matches the theoretical expectation perfectly, which verifies our theoretical analysis.

\section{The Proof of Theorem~\ref{theorem: estimator}}
\label{appendix: the proof of theorem estimator}
Based on the lemma above, we prove the unbiaseness and the error bound of the estimator. 
\begin{proof}
    We first prove the unbiasedness.
    When $\mathbf{o}$ and $\mathbf{q}$ are collinear, the unbiasedness can be trivially verified by definition. 
    When $\mathbf{o}$ and $\mathbf{q}$ are non-collinear, in order to prove the unbiasedness, it suffices to prove that the error term of the estimator in (\ref{eq: simplify}) equals to 0 in expectation. Letting $\mathbf{e}_1:= \frac{\mathbf{q}  - \left< \mathbf{q},\mathbf{o}  \right> \mathbf{o}}{\| \mathbf{q}  - \left< \mathbf{q},\mathbf{o}  \right> \mathbf{o}\|}$, we deduce from $\mathbb{E}\left[ \frac{\left< \mathbf{\bar o}, \mathbf{e}_1 \right> }{ \left< \mathbf{\bar o}, \mathbf{o}  \right> }  \right] $ as follows.
    \begin{align}
         \mathbb{E} \left[ \frac{\left< \mathbf{\bar o}, \mathbf{e}_1  \right> }{ \left< \mathbf{\bar o}, \mathbf{o}  \right> }  \right] = & \mathbb{E} \left[ \sqrt {1- \left< \mathbf{\bar o}, \mathbf{o}  \right> ^2 } \cdot X_1 / \left< \mathbf{\bar o}, \mathbf{o}  \right>    \right] \label{eq: theorem1 by lemma}
        \\= &\mathbb{E} \left[ \sqrt {1- \left< \mathbf{\bar o}, \mathbf{o}  \right> ^2 } / \left< \mathbf{\bar o}, \mathbf{o}  \right>    \right] \cdot \mathbb{E} \left[ X_1 \right]  \label{eq: theorem1 by independency}
        \\= &\mathbb{E} \left[ \sqrt {1- \left< \mathbf{\bar o}, \mathbf{o}  \right> ^2 } / \left< \mathbf{\bar o}, \mathbf{o}  \right>    \right] \cdot 0 = 0
        \label{eq: theorem1 by distribution}
    \end{align}
    where (\ref{eq: theorem1 by lemma}) is by Lemma~\ref{lemma: joint distribution}. (\ref{eq: theorem1 by independency}) is due to the independence between $\left< \mathbf{\bar o}, \mathbf{o}  \right> $ and $X_1$. (\ref{eq: theorem1 by distribution}) is because the distribution of $X_1$ (i.e., $p_{D-1}$) has the mean of 0. Finally, based on (\ref{eq: simplify}), we finish the proof of the unbiasedness.
    
    We then prove the error bound. When $\mathbf{o}$ and $\mathbf{q}$ are collinear, the error is zero as is specified by Section~\ref{subsubsec: construct unbiased estimator}.
    We prove the error bound for the non-collinear case as follows.
    \begin{align}
        &\mathbb{P} \left\{ \left| \frac{\left< \mathbf{\bar o}, \mathbf{q}  \right> }{\left< \mathbf{\bar o,\mathbf{o} } \right> } -\left< \mathbf{o,q} \right>   \right| >  \sqrt{\frac{{1 - \left< \mathbf{\bar o}, \mathbf{o}  \right>^2}}{\left< \mathbf{\bar o}, \mathbf{o}  \right>^2 }} \cdot \frac{\epsilon_0}{\sqrt {D-1} }   \right\}  \\
        =& \mathbb{P} \left\{ \sqrt {1 - \left< \mathbf{o,q} \right>^2 } 
        \left|  \frac{\left< \mathbf{\bar o}, \mathbf{e}_1  \right> }{\left< \mathbf{\bar o}, \mathbf{o}  \right> }   \right|
        >  \sqrt{\frac{{1 - \left< \mathbf{\bar o}, \mathbf{o}  \right>^2}}{\left< \mathbf{\bar o}, \mathbf{o}  \right>^2 }} \cdot \frac{\epsilon_0}{\sqrt {D-1} }   \right\} \label{eq: theorem 3.3 plugin lemma3.1}\\
        \le& \mathbb{P} \left\{ 
        \left|  \frac{\left< \mathbf{\bar o}, \mathbf{e}_1  \right> }{\left< \mathbf{\bar o}, \mathbf{o}  \right> }   \right|
        >  \sqrt{\frac{{1 - \left< \mathbf{\bar o}, \mathbf{o}  \right>^2}}{\left< \mathbf{\bar o}, \mathbf{o}  \right>^2 }} \cdot \frac{\epsilon_0}{\sqrt {D-1} }   \right\} \label{eq: theorem 3.3 simple relaxation}\\
        =& \mathbb{P} \left\{ \sqrt{\frac{{1 - \left< \mathbf{\bar o}, \mathbf{o}  \right>^2}}{\left< \mathbf{\bar o}, \mathbf{o}  \right>^2 }} \cdot 
        \left|  X_1  \right|
        >  \sqrt{\frac{{1 - \left< \mathbf{\bar o}, \mathbf{o}  \right>^2}}{\left< \mathbf{\bar o}, \mathbf{o}  \right>^2 }}\cdot \frac{\epsilon_0}{\sqrt {D-1} }   \right\} \label{eq: theorem 3.3 plugin lemma3.2} \\
        =& \mathbb{P} \left\{ 
        \left|  X_1  \right|
        >  \frac{\epsilon_0}{\sqrt {D-1} }   \right\} \le 2e^{-c_0 \epsilon_0^2}\label{eq: theorem 3.3 plugin concentration}
    \end{align}
    where (\ref{eq: theorem 3.3 plugin lemma3.1}) is by Lemma~\ref{lemma: geometry}. (\ref{eq: theorem 3.3 simple relaxation}) relaxes $\sqrt {1 -  {\left< \mathbf{o,q} \right>^2 } } $ to 1. (\ref{eq: theorem 3.3 plugin lemma3.2}) is due to Lemma~\ref{lemma: joint distribution}. (\ref{eq: theorem 3.3 plugin concentration}) applies Lemma~\ref{lemma: distribution}.
\end{proof}

\begin{table*}[t]
\caption{The Summary of the Analysis for $B_q$.}
\begin{tabular}{c|cc|cc|c}
\label{table: BQ error}
        & Error & $\Delta$ & Overall Error        & Target & $B_q$ \\ \hline
Trivial & $O(\sqrt {D} \cdot \Delta)$  &  $O(1/2^{B_q})$                             & $O(\sqrt {D}/2^{B_q})$ &  $O(1/\sqrt{D})$  & $\Theta(\log D)$      \\ \hline
Ours    & $O(\Delta)$  &  $O(\sqrt {\frac{\log D}{D}} / 2^{B_q})$    & $O(\sqrt {\frac{\log D}{D}} / 2^{B_q})$ &  $O(1/\sqrt{D})$   & $\Theta(\log \log D)$
\end{tabular}
\end{table*}
\section{The Analysis for $B_q$}
\label{appendix: analysis for BQ}
In this section, we prove that $B_q=\Theta(\log \log D)$ suffices to guarantee that the error introduced by the uniform scalar quantization is much smaller than the error of the estimator itself, i.e., $O(1/\sqrt {D})$.
We prove the statement in two steps.
\underline{First}, we will prove that the error introduced by the uniform scalar quantization is $O(\Delta)$ in Section~\ref{appendix, subsec error of randomized uniform scalar quantization} (recall that $\Delta= \left( \max \mathbf{q'}[i]- \min \mathbf{q'}[i]  \right) / (2^{B_q}-1)$). Specifically, recall that we approximately compute $\left< \mathbf{\bar x}, \mathbf{q}'  \right> $ as $\left< \mathbf{\bar x}, \mathbf{\bar q}  \right> $. Then the error equals to $\left| \left< \mathbf{\bar x}, \mathbf{\bar q}-\mathbf{q}'  \right> \right| $. Note that $\mathbf{\bar x}$ has its entries of $\pm 1/\sqrt {D}$ and $\mathbf{\bar q}[i]-\mathbf{q}'[i]\in [-\Delta,\Delta]$. Then a trivial bound by the triangle's inequality yields 
$\left| \left< \mathbf{\bar x}, \mathbf{\bar q}-\mathbf{q}'  \right> \right| \le D \cdot \frac{1}{\sqrt {D} } \cdot \Delta = \sqrt {D} \cdot \Delta$.
We note that the bound is weak because it does not consider the cancellation of under-estimate and over-estimate. 
We will prove a stronger probabilistic bound of $O(\Delta)$ by considering the randomized uniform scalar quantization algorithm presented in Section~\ref{subsubsec: quantize query vector} and quantitatively analyzing the cancellation.
\underline{Second}, we will prove that $\Delta = O(\sqrt {\frac{\log D}{D}} / 2^{B_q})$ in Section~\ref{appendix, delta}. 
Recall that $\mathbf{q}'=P^{-1} \mathbf{q} $ is a random vector which follows the uniform distribution on the unit sphere. A trivial bound gives that $\left( \max \mathbf{q'}[i]- \min \mathbf{q'}[i]  \right) = O(1)$. 
We note that the bound is weak because it does not consider the concentration phenomenon of the entries of $\mathbf{q'}$ (see Lemma~\ref{lemma: distribution}). 
We will prove a stronger probabilistic bound 
by quantitatively considering the concentration phenomenon.

Based on the two probabilistic bounds, we derive the overall error by applying the union bound. The conclusions are summarized in Table~\ref{table: BQ error}.
Making the overall error be $O(1/\sqrt {D})$, we have $B_q = \Theta(\log \log D)$. We note that based on the trivial bounds, it is necessary to set $B_q = \Theta (\log D)$ to guarantee the same error bound, which is exponentially worse than our result. 

\subsection{The Error of Randomized Uniform Scalar Quantization}
\label{appendix, subsec error of randomized uniform scalar quantization}
As is discussed above, the error introduced by the uniform scalar quantization is 
\begin{align}
    \left| \left< \mathbf{\bar x}, \mathbf{\bar q}-\mathbf{q}'   \right>  \right| =\left| \sum_{i=1}^{D}  \mathbf{\bar x}[i] \cdot (\mathbf{\bar q}[i]-\mathbf{q}'[i])    \right|
\end{align}
Due to the randomized uniform scalar quantization presented in Section~\ref{subsubsec: quantize query vector}, each term of the error, i.e., $\mathbf{\bar x}[i] \cdot (\mathbf{\bar q}[i]-\mathbf{q}'[i])$, is a random variable. 
The $D$ random variables are independent to each other. Each term has the expected value of 0 (see Section~\ref{subsubsec: quantize query vector}) and has their values bounded by $[-\Delta /\sqrt {D}, +\Delta /\sqrt {D}]$. Now the question is to analyze the summation of $D$ such random variables. 
We note that the Hoeffding's inequality immediately answers the question~\cite{vershynin_2018}. We restate the inequality in the following lemma.

\begin{lemma}[Hoeffding's Inequality~\cite{vershynin_2018}]
Let $X_1,...,X_n$ be independent random variables, such that $X_i \in [a_i,b_i], \forall 1\le i\le n$. Let $S_n=\sum_{i=1}^{n} X_i$. Then
\begin{align}
    \mathbb{P} \left\{ \left| S_n - \mathbb{E}\left[ S_n \right]   \right| \ge t  \right\}   \le 2\exp \left( - \frac{2t^2}{\sum_{i=1}^{n} (b_i-a_i)^2}  \right) 
\end{align}
\end{lemma}

In our case, we note that $a_i=-\Delta /\sqrt {D}$, $b_i=+\Delta /\sqrt {D}$.  $\mathbb{E}\left[ S_n \right] =\mathbb{E}\left[ \sum_{i=1}^{n} X_i \right]=\sum_{i=1}^{n}  \mathbb{E}\left[ X_i \right] = 0$. 
It immediately yields the following conclusion.
\begin{align}
    \mathbb{P} \left\{ \left| \sum_{i=1}^{D}  \mathbf{\bar x}[i] \cdot (\mathbf{\bar q}[i]-\mathbf{q}'[i])    \right|\ge t  \right\}  \le 2\exp \left( - \frac{t^2}{2\Delta^2}  \right) 
    \\ \mathbb{P} \left\{ \left| \sum_{i=1}^{D}  \mathbf{\bar x}[i] \cdot (\mathbf{\bar q}[i]-\mathbf{q}'[i])    \right|\ge \Delta u  \right\}  \le 2\exp \left( - \frac{u^2}{2}  \right) \label{appendix: eq 65}
\end{align}
where (\ref{appendix: eq 65}) is by letting $u=t/\Delta$. The conclusion shows that the error is bounded by $O(\Delta)$ with high probability. 

\subsection{The Analysis for $\Delta$}
\label{appendix, delta}
Next we prove $\Delta = O(\sqrt {\frac{\log D}{D} } /2^{B_q})$ with high probability. 
Recall that $\Delta= \left( \max_{1\le i\le D} \mathbf{q'}[i]- \min_{1\le i\le D} \mathbf{q'}[i]  \right) / (2^{B_q}-1)$.
Note that $\max_{1\le i\le D}(\mathbf{q'}[i])- \min_{1\le i\le D} (\mathbf{q'}[i]) \le 2 \max_{1\le i\le D} |\mathbf{q'}[i]|$.
In order to prove the original statement, it suffices to prove that $\max_{1\le i\le D} |\mathbf{q'}[i]| = O(\sqrt {\frac{\log D}{D}})$ with high probability, which we prove as follows.
    \begin{align}
        &\mathbb{P} \left\{ \max_{1 \le i\le D} |\mathbf{q'}[i]| \ge \sqrt {\frac{\log D + t}{c_0 D} }  \right\}  
        \\ = & \mathbb{P} \left\{  \exists 1\le i \le D, |\mathbf{q'}[i]| \ge \sqrt {\frac{\log D + t}{c_0 D} }  \right\} 
        \\ \le & D \cdot \mathbb{P} \left\{ |\mathbf{q'}[1]| \ge \sqrt {\frac{\log D + t}{c_0 D} }  \right\}  \label{eq 69: union bound}
        \\ \le & 2 \exp \left( -c_0 \cdot  \frac{\log D + t}{c_0} + \log D \right) =2 \exp (-t) \label{eq 70: by distribution}
    \end{align}
    where (\ref{eq 69: union bound}) is by union bound. (\ref{eq 70: by distribution}) is by Lemma~\ref{lemma: distribution}.
The conclusion shows that $\Delta = O (\sqrt {\frac{\log D}{D} }/2^{B_q})$ with high probability.

\section{Discussion on the Normalization}
\label{subsubsec: normalization}
We note that our current theoretical analysis (without any assumptions on the data) provides an additive error bound like \cite{2017_focs_additive_error}. 
Let $dist'^2$ be the estimated squared distance based on our estimator, where $dist'^2=\| \mathbf{o}_r-\mathbf{c} \|^2 + \| \mathbf{q}_r-\mathbf{c} \|^2 -2 \cdot \|\mathbf{o}_r-\mathbf{c}\| \cdot \|\mathbf{q}_r-\mathbf{c}\|  \cdot \frac{\left< \mathbf{\bar o}, \mathbf{q}  \right> }{\left< \mathbf{\bar o}, \mathbf{o}  \right>}  $.
Considering that we do not normalize the dataset with the centroids but with the origin of the space, then based on the Equation (\ref{eq: concise presentation of error bound}) in Theorem~\ref{theorem: estimator}, we immediately have
\begin{align}
    \left| dist'^2-\| \mathbf{o}_r-\mathbf{q}_r\|^2   \right| = \|\mathbf{o}_r\| \cdot \|\mathbf{q}_r\|  \cdot O\left( \frac{1}{\sqrt {D} } \right) \ w.h.p. \label{eq: 71}
\end{align}
where w.h.p. is short for ``with high probability''.

When the data vectors are well normalized and spread evenly on the unit hypersphere, 
we assume that 
the data vector after normalization, i.e., $\mathbf{o} = \frac{\mathbf{o}_r-\mathbf{c}}{\| \mathbf{o}_r-\mathbf{c}\|} $, 
follows the uniform distribution on the unit hypersphere. Then we can derive the following corollary which presents a multiplicative error bound like \cite{SIAM_JOC_2022_indyk}. 
\begin{corollary}
    Assuming that $\mathbf{o}$ follows the uniform distribution on the unit  hypersphere. We have 
    \begin{align}
        \left| \frac{dist'^2-\| \mathbf{o}_r-\mathbf{q}_r\|^2}{\| \mathbf{o}_r-\mathbf{q}_r\|^2}  \right| = O\left( \frac{1}{\sqrt {D} } \right) \ w.h.p.
    \end{align}
\end{corollary}
\begin{proof}
    \begin{align}
        &\left| \frac{dist'^2 - \| \mathbf{q}_{r} - \mathbf{o}_r\|^2}{\| \mathbf{q}_{r} - \mathbf{o}_r\|^2} \right| =  \left| \frac{2\| \mathbf{q}_r-\mathbf{c}\|\| \mathbf{o}_r-\mathbf{c}\|  }{\| \mathbf{q}_{r} - \mathbf{o}_r\|^2} \right| \cdot O\left( \frac{1}{\sqrt {D} } \right)  \label{eq: 73}
\\=& \left| \frac{2\| \mathbf{q}_r-\mathbf{c}\|\| \mathbf{o}_r-\mathbf{c}\|   }{\| \mathbf{q}_r-\mathbf{c}\|^2 + \| \mathbf{o}_r-\mathbf{c}\|^2 -2 \left< \mathbf{o}_r-\mathbf{c}, \mathbf{q}_r-\mathbf{c}\right> } \right| \cdot O \left( \frac{1}{\sqrt {D} } \right) \label{eq: 74}
\\=& \left| \frac{1}{\frac{1}{2} \left( \frac{\| \mathbf{q}_r-\mathbf{c}\|}{\| \mathbf{o}_r-\mathbf{c}\|}   + \frac{\| \mathbf{o}_r-\mathbf{c}\|}{\| \mathbf{q}_r-\mathbf{c}\|}\right) - \left< \mathbf{o}, \mathbf{q}\right>} \right| \cdot O \left( \frac{1}{\sqrt {D} } \right)  \label{eq: 75}
\\ \le& \left| \frac{1}{1 - \left< \mathbf{o}, \mathbf{q}\right>} \right| \cdot O \left( \frac{1}{\sqrt {D} } \right)  = O \left( \frac{1}{\sqrt {D} } \right) \ w.h.p. \label{eq: 76}
    \end{align}
    where (\ref{eq: 73}) is due to (\ref{eq: 71}). (\ref{eq: 74}) is by elementary linear algebra. (\ref{eq: 75}) simplifies (\ref{eq: 74}). The first inequality of (\ref{eq: 76}) is due to the numerical inequality that $\left( x+\frac{1}{x} \right) \ge 2, \forall x>0$. The second equality of (\ref{eq: 76}) holds because $\mathbf{o}$ follows the uniform distribution on the unit hypersphere. In particular, due to the concentration inequality (Lemma~\ref{lemma: distribution}), $\left< \mathbf{o},\mathbf{q}  \right> $ is highly concentrated around 0, e.g., it has sufficiently high probability to be upper bounded by $\frac{1}{2}$. 
\end{proof}
We emphasize that the corollary is not fully rigorous as it depends on the assumption that $\mathbf{o}$ follows the uniform distribution on the unit hypersphere. Note that all our other theoretical results do not rely on any assumptions on the data, i.e., the additive bound holds regardless of the data distribution. 
In the present work, we only adopt a simple and natural method of normalization (i.e., with the centroids of IVF) to instantiate our scheme of quantization, while we have yet to extensively explore the normalization step itself. We shall leave it as future work to thoroughly and rigorously study the normalization problem.

To verify the effectiveness of the current method of normalization, we measure the empirical uniformity of the normalized data vectors with the entropy of each bit in our quantization codes~\footnote{In our quantization codes, let $p$ be the proportion of 1's in the $i$th bit of the whole dataset. Then the entropy of the $i$th bit equals to $-[p\log_2 (p) + (1-p)\log_2 (1-p)]$. We report the summation of the entropy over all the bits. The result is normalized by the length of the quantization codes for the ease of study.}. The larger the entropy is, the more uniform the normalized dataset is (and at the same time, the more informative the quantization codes are). 
We report that on all the datasets, the entropy is over 99.9\% of the length of the quantization codes (in other word, for every bit in the quantization codes, the number of 1's is almost the same as the number of 0's), indicating that normalizing the dataset with the centroids of IVF is empirically effective.

\section{Discussion on the Ablation Study}
\label{appendix: ablation}
\begin{figure}[thb]
    \vspace{-3mm}
    \centering
    \includegraphics[width=0.8\linewidth]{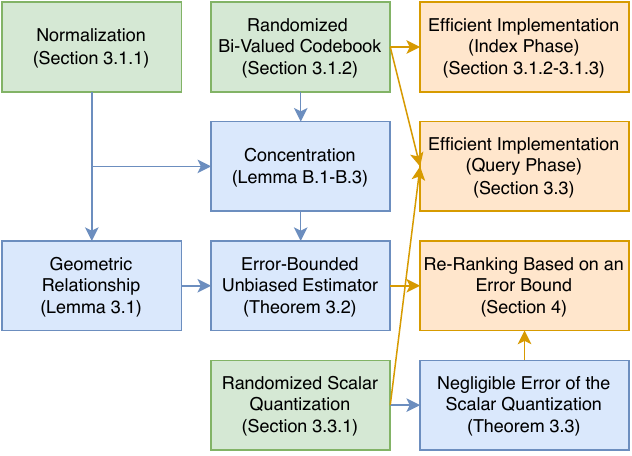}
    \vspace{-2mm}
    \caption{Dependency among Algorithm Components, Theoretical Conclusions, and Efficient Implementations}.
    \label{figure: workflow}
    \vspace{-4mm}
\end{figure}

We emphasize that RaBitQ is a method with rigorous theoretical guarantee. The newly proposed codebook construction and distance estimation are an integral whole. 
The ablation of any component will cause the loss of the theoretical guarantee (i.e., the performance is no more theoretically predictable) and further disable the error-bound-based re-ranking for ANN (Section~\ref{sec: RaBitQ for ANN}). We summarize the dependency of the theoretical conclusions and efficient implementations on our algorithm components in detail
in Figure~\ref{figure: workflow}. The green, blue and orange boxes represent the components in our algorithm, the theoretical conclusions and the implementations respectively. 
An arrow indicates that a particular component or theoretical conclusion is used for achieving the downstream conclusion or implementation. 
In addition, we would also like to highlight that although RaBitQ achieves the asymptotic optimality in terms of the worst-case error bound, for datasets which have certain promising properties, 
it is still possible to obtain better empirical performance via further (possibly heuristic) optimization.
We believe this is an interesting research topic and we would like to leave it as future work.

Despite that the ablation is irrational in view of theory, we provide several empirical studies to discuss how each component empirically affects the performance~\footnote{Recall that the error-bound-based re-ranking would be disabled when any of the components is ablated. Using it for ANN entails exhaustive tuning of the parameters of re-ranking, which would heavily affect the results of ANN search.
Thus, the results of ANN would be less insightful for an ablation study. We only report the results of distance estimation.}. 
We investigate the effects of removing a certain component from RaBitQ while keeping the others.

\begin{figure*}[thb]
  \centering 
    \includegraphics[width=13cm]{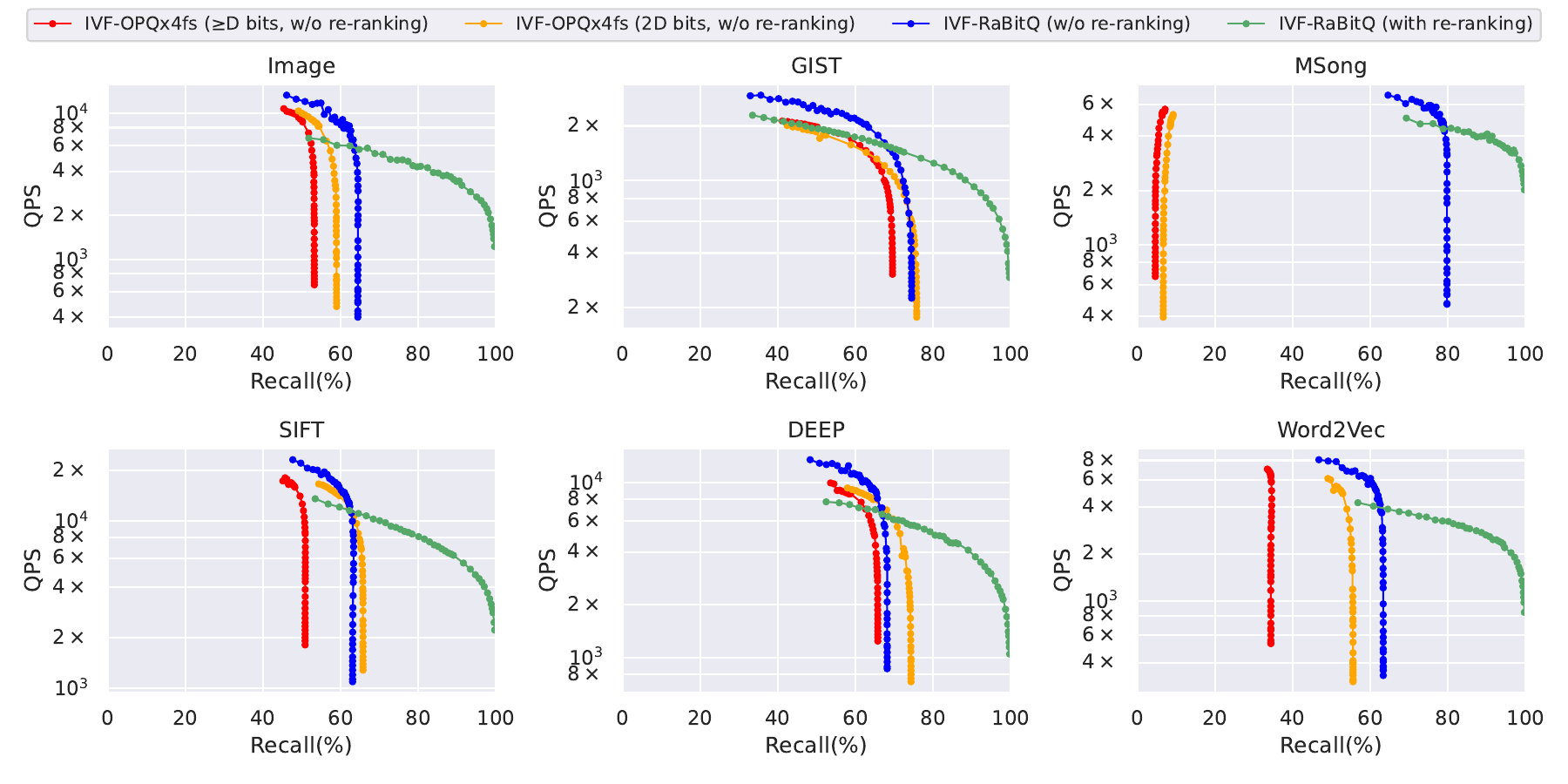}
  \vspace*{-4mm}
  \caption{Time-Accuracy Trade-Off for ANN Search (with and w/o re-ranking).}
  \vspace*{-4mm}
  \label{figure:wo_reranking}
\end{figure*}

\subsection{The Ablation of the Codebook Construction}
\label{subsec:ablation-codebook}
Recall that in Section~\ref{subsec: index}, for quantizing a set of normalized data vectors, we construct a quantization codebook by randomly rotating a set of bi-valued vectors.
We consider replacing our randomized codebook with a learned heuristic codebook of PQ while keeping other components. 
We note that with our randomized codebook, RaBitQ has rigorous theoretical guarantee and achieves asymptotic optimality. However, with a learned heuristic codebook, the performance of the method will be no more theoretically predictable.
Though it is often intuitive to expect that replacing a randomized algorithm with a learned algorithm may improve the performance, in RaBitQ, whose design is an integral whole, a learned codebook could be less suitable than a randomized codebook.
Table~\ref{table:accuracy ablation} presents the average relative error and maximum relative error of the ablation study on the GIST dataset. It clearly shows that replacing a randomized codebook with a learned one degrades both the general quality and the robustness of distance estimation.
The results imply that despite that PQ, as a learning-based method, has a large search space of codebooks, the heuristic learning process (which is based on a heuristic objective function and an approximate optimization algorithm) only finds a suboptimal solution among the search space. 
Similar findings can be observed in the main experiments (Section~\ref{subsubsec: time-accuracy trade-off per vector}), i.e., LSQ has its search space of codebooks even larger than PQ, yet it has worse performance in most cases.

\begin{table}[h]
\vspace{-2mm}
\caption{Ablation Study of the Codebook Construction (GIST).}
\label{table:accuracy ablation}
\vspace{-4mm}
\begin{tabular}{c|c|c}
\hline
 &  Ave. Rel. Error (\%)      & Max. Rel. Error (\%)   \\ \hline
Rand. Codebook & 1.675 & 13.043 \\ \hline
Learned Codebook & 3.049 & 34.375 \\ \hline
\end{tabular}
\vspace{-4mm}
\end{table}

\subsection{The Ablation of the Estimator}
\label{subsec:ablation-estimator}
\begin{figure}[bht]
    \vspace{-2mm}
    \centering
    \includegraphics[width=0.5\linewidth]{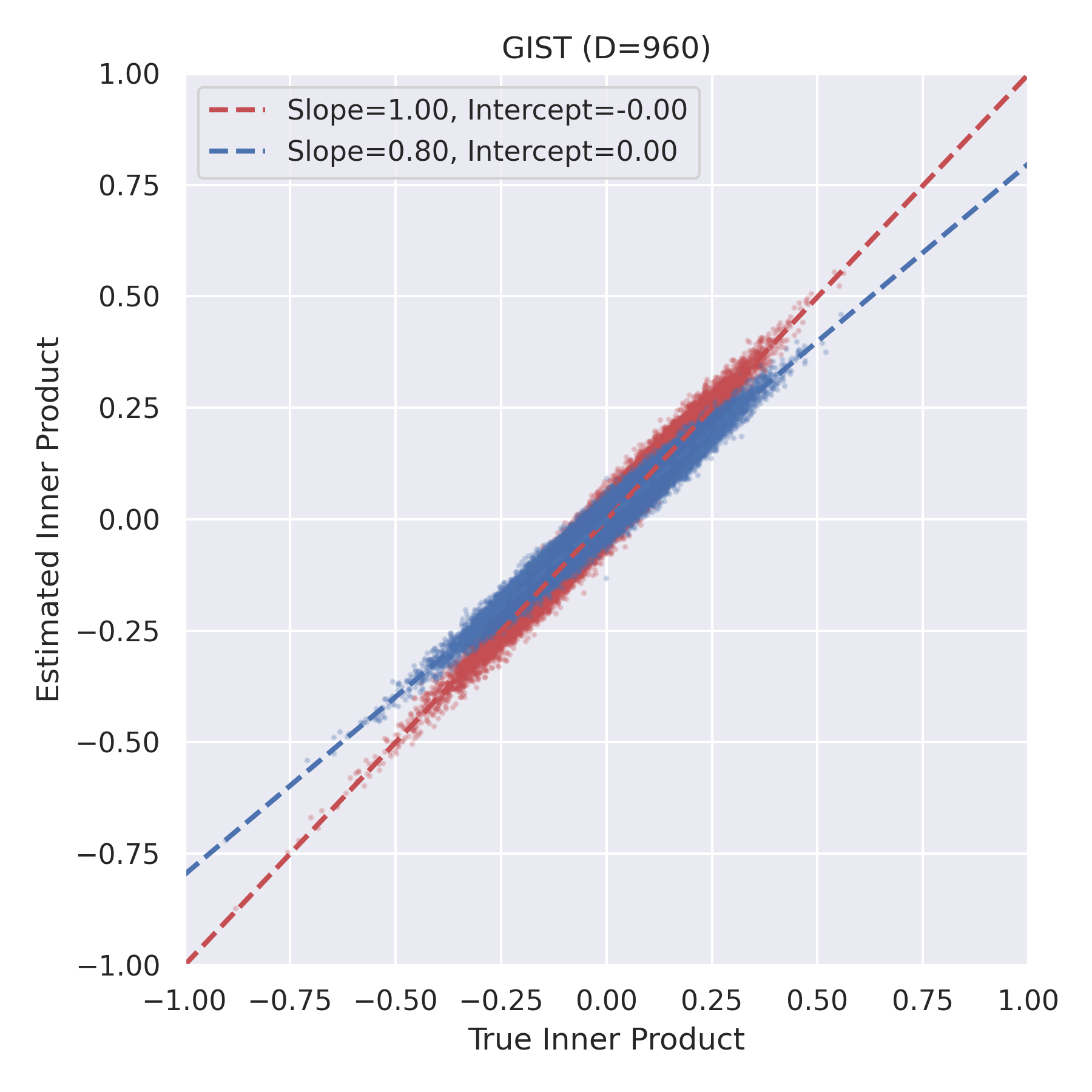}
    \vspace{-4mm}
    \caption{ Ablation Study of the Estimator. The red points represent the results based on the estimator $\frac{\left< \mathbf{\bar o}, \mathbf{q}  \right> }{\left< \mathbf{\bar o}, \mathbf{o}  \right>} $. The blue points  represent the results based on the estimator $\left< \mathbf{\bar o}, \mathbf{q}  \right>$.}
    \label{fig: ablation study of the estimator}
    \vspace{-4mm}
\end{figure}

Recall that as is discussed in Section~\ref{subsec: unbiased estimator}, unlike PQ which simply treats the quantized data vector as the data vector, our method explicitly analyzes the geometric relationship between the vectors and constructs an unbiased estimator accordingly. In this part, we ablate our estimator and adopt an alternative estimator $\left< \mathbf{\bar o}, \mathbf{q}  \right> $ by treating the quantized data vector as the data vector as PQ does. 
In particular, we collect $10^7$ pairs of the estimated inner products and the true inner products from the first 10 query vectors and the $10^6$ data vectors of the GIST dataset. 
Figure~\ref{fig: ablation study of the estimator} shows the scatter plots of the true inner product and the estimated inner product. The red points represent the results based our unbiased estimator $\frac{\left< \mathbf{\bar o}, \mathbf{q}  \right> }{\left< \mathbf{\bar o}, \mathbf{o}  \right>} $. The blue points represent the results based on the estimator $\left< \mathbf{\bar o}, \mathbf{q}  \right> $. We fit the two set of points with linear regression. The results clearly show that our estimator is unbiased while the estimator $\left< \mathbf{\bar o}, \mathbf{q}  \right> $ is biased by a ratio of around 0.8.
Table~\ref{table:estimator ablation} further presents the average relative error and maximum relative error of the estimated distances on the GIST dataset. 
It clearly shows that taking $\left< \mathbf{\bar o}, \mathbf{q}  \right> $ as the estimator degrades both the general quality and the robustness of the distance estimation.
Moreover, we note that based on the estimator $\left< \mathbf{\bar o}, \mathbf{q}  \right> $, the original theoretical error bound (Theorem~\ref{theorem: estimator}) does not hold anymore. Thus, it is inapplicable for the error-bound-based re-ranking for in-memory ANN search (Section~\ref{sec: RaBitQ for ANN}).

\begin{table}[h]
\vspace{-2mm}
\caption{Ablation Study of the Estimator (GIST).}
\label{table:estimator ablation}
\vspace{-4mm}
\begin{tabular}{c|c|c}
\hline
 & Ave. Rel. Error (\%)      & Max. Rel. Error (\%)   \\ \hline
$\frac{\left<\mathbf{\bar o}, \mathbf{q}\right>}{\left<\mathbf{\bar o}, \mathbf{o}\right>}    $ & 1.675 & 13.043 \\ \hline
$\left< \mathbf{\bar o}, \mathbf{q}  \right> $& 2.196 & 52.400 \\ \hline
\end{tabular}
\vspace{-4mm}
\end{table}

\subsection{The Ablation of the Re-Ranking}
\label{appendix: without reranking}
As is discussed in Section~\ref{sec: RaBitQ for ANN}, despite that RaBitQ provides the guarantee on the distance estimation, when the distances (to the query) of two different data vectors are extremely close to each other, the guaranteed accuracy might be insufficient for ranking them correctly.
Re-ranking, in this case, is necessary for producing high recall. Figure~\ref{figure:wo_reranking} plots the ``QPS''-``recall'' curves of RaBitQ with and without re-ranking. It shows that re-ranking is indeed necessary for achieving the robust performance of ANN search.

\end{document}